\documentclass[sigconf]{acmart}

\copyrightyear{2026}
\acmYear{2026}
\setcopyright{cc}
\setcctype{by}
\acmConference[aiDM '26]{Nineth International Workshop on Exploiting Artificial Intelligence Techniques for Data Management }{May 31-June 05, 2026}{Bengaluru, India}
\acmBooktitle{Nineth International Workshop on Exploiting Artificial Intelligence Techniques for Data Management (aiDM '26), May 31-June 05, 2026, Bengaluru, India}
\acmDOI{10.1145/3814940.3815329}
\acmISBN{979-8-4007-2719-1/2026/05}

\usepackage{graphicx}
\usepackage{subcaption}
\usepackage{multirow}
\usepackage{subcaption}
\usepackage{placeins}
\usepackage{hyperref}
\usepackage{hyperxmp}
\usepackage[htt]{hyphenat}

\usepackage{cuted}

\usepackage{listings}
\usepackage{xcolor}
\usepackage{verbatim}
\usepackage{url}

\definecolor{codegreen}{rgb}{0,0.6,0}
\definecolor{codegray}{rgb}{0.5,0.5,0.5}
\definecolor{codepurple}{rgb}{0.58,0,0.82}
\definecolor{backcolour}{rgb}{0.95,0.95,0.92}

\lstdefinestyle{mystyle}{
    backgroundcolor=\color{backcolour},   
    commentstyle=\color{codegreen},
    keywordstyle=\color{magenta},
    numberstyle=\tiny\color{codegray},
    stringstyle=\color{codepurple},
    basicstyle=\ttfamily\footnotesize,
    breakatwhitespace=false,         
    breaklines=true,                 
    captionpos=b,                    
    keepspaces=true,                 
    numbers=left,                    
    numbersep=5pt,                  
    showspaces=false,                
    showstringspaces=false,
    showtabs=false,                  
    tabsize=2
}

\usepackage{soul}
\usepackage{todonotes}
\presetkeys{todonotes}{inline}{}

\title{Same Data, Different Schemas: Robustness of LLM-based Text-to-SQL}
\title{Same Data, Different Schemas: Robustness of LLM-based Text-to-SQL}
\author{Nitin Kanchinadam, Aditya Menachery, Amol Deshpande}
\email{nkanchin@umd.edu, adityam@umd.edu, amol@umd.edu}
\affiliation{%
\institution{University of Maryland, College Park, USA\\[20pt]}
}

\begin{document}

\begin{abstract}
Large language models (LLMs) consistently achieve strong results on text-to-SQL benchmarks, %translating
but their robustness to schema variations
remains poorly understood. Recent work suggests that the schema structure matters, but does not provide a clear and systematic way to evaluate model behavior when different schemas represent the same underlying data.
We address this problem by presenting a framework to evaluate and benchmark text-to-SQL techniques over equivalent relational schemas generated from a common E/R model. By varying the ``shredding'' choices used to translate the conceptual design into relations, we create multiple schema variants that differ structurally while preserving the same underlying semantics. This gives us a controlled setting in which the natural language questions and data remain fixed, and only the schema changes.
We use this framework to evaluate four leading LLMs on the same questions across multiple schema variants
(for two separate domains), and  %, comparing both the generated SQL and the execution results. 
summarize consistency patterns using pairwise comparison heatmaps. Our results show that schema structure
significantly affects LLM behavior: across conceptually equivalent schemas, models often produce SQL queries
with very different answers. We also find that providing additional context (specifically, the original E/R specification) improves the performance somewhat, but does not fully ameliorate the inconsistencies.
In addition to demonstrating that the current text-to-SQL evaluations miss an important notion of robustness, our framework provides a way to generate a large number of synthetic datasets that can be used to train new models, and suggests a mechanism to make text-to-SQL more robust by generating additional candidate plans for a given natural language query through systematic schema variations.
\end{abstract}


\maketitle

\section{Introduction}

Text-to-SQL (NL2SQL) systems aim to bridge the gap between natural language and structured query languages, and there is
a large body of work on developing such
systems~\cite{wang2020rat,katsogiannis2023survey,pourreza2023din,pourreza2024chase,li2024codes,maamari2024death,liu2024survey,li2023can,hong2025next,bhaskar2023benchmarking,ding2025ambisql}. 
Recent advances in large language models (LLMs) have led to major improvements on standard benchmarks like Spider~\cite{yu2018spider} and BIRD~\cite{livesqlbench2025}. On Spider 2.0, leading systems now report test accuracy in the mid-80s to low-90s, and prompt-based systems using strong LLMs have reached 96.7\% on the test set\footnote{https://spider2-sql.github.io, Retrieved May 8, 2026}.
At the same time, anecdotal evidence from real-world deployments indicates a large gap between success on
standard benchmarks and performance in more realistic settings~\cite{liu2025nl2sql}, leading to continued
development of new techniques~\cite{shi2024surveyemployinglargelanguage,hong2025next} as well as benchmarks (e.g., BIRD-Interact~\cite{huo2026birdinteract}).

A common theme in the recent work is that schema structure matters. Many successful systems explicitly model schema structure, perform schema linking, or both, and prior work has argued that schema linking is one of the main bottlenecks in Text-to-SQL~\cite{lei2020re}. More recent work has also shown that current models can be sensitive to changes in schema organization. Li et al., show that the same question may require substantially different SQL when the schema structure changes, and that model performance drops sharply on schema-altered versions of existing benchmarks~\cite{li2023exploring}. Fürst et al., study robustness across three different data models for the same football domain and show that performance can vary significantly across alternative schemas built over the same underlying data~\cite{furst2024evaluating}. Other recent work has looked at robustness under schema evolution and at the effect of normalization choices, again showing that schema design can have a strong effect on LLM-based SQL generation~\cite{zhang2026evoschema,kohita2025exploring}.

\begin{figure*}[t]
    \centering
    \includegraphics[width=\textwidth]{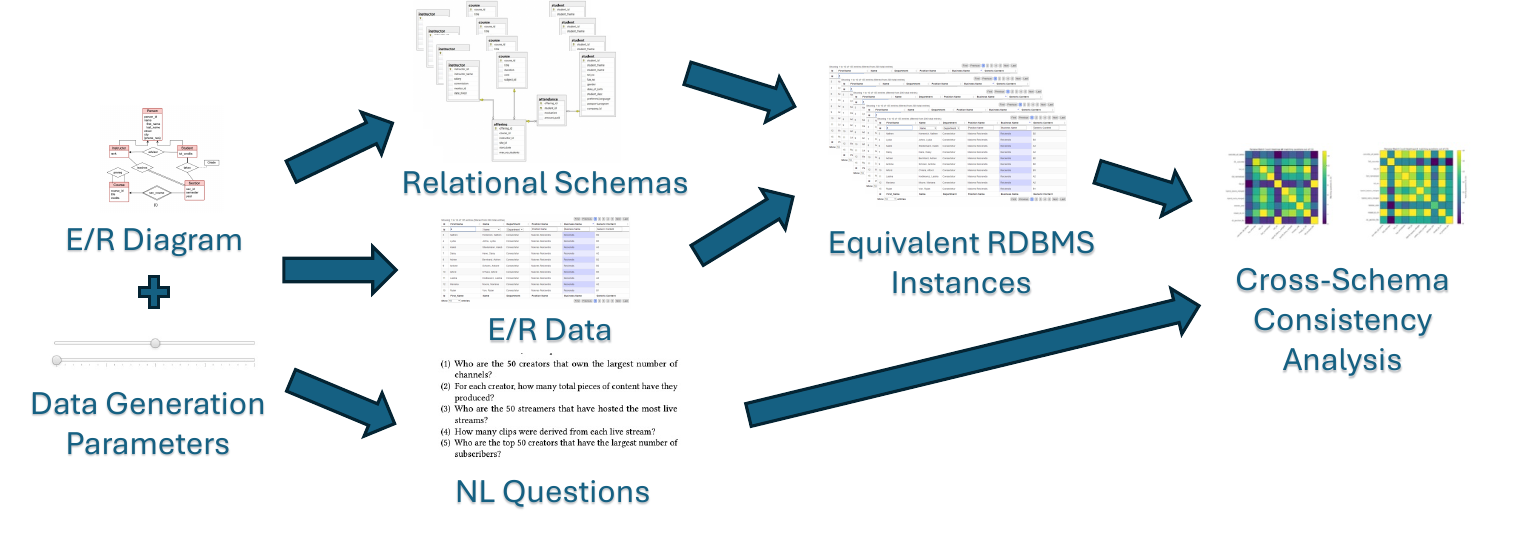}
    \vspace{-15pt}
    \caption{Overall benchmark generation and evaluation pipeline. Most of the steps, including initial E/R diagram, can themselves be fully automated using LLMs.}
    \vspace{-5pt}
    \label{fig:dataset_pipeline}
\end{figure*}

These results point to an important gap in current evaluation. Standard text-to-SQL benchmarks mostly test cross-domain generalization: can a model answer questions over databases and schemas it has not seen before? %We study a somewhat different problem in this work. 
However, in many real applications, the same underlying data can be represented by different relational schemas. This can happen because of normalization and denormalization choices, because relationships are represented
differently, or because an E/R design is translated into relations in different ways. This is especially common in domains with inheritance or subtype structure, where there are multiple reasonable shredding choices. From the
perspective of the user, however, the questions or the data have not changed, only the logical representation of the data in the database, i.e., the schema, has changed.

This leads to the question we study in this paper: \textit{Do LLM-based Text-to-SQL systems produce consistent and correct SQL when the schema changes, even when the question and the underlying data remain the same?}

We study this question through the lens of alternative E/R-to-relational shredding choices. Starting from a common E/R model, we generate multiple relational schemas that are
different at the logical level but equivalent at the conceptual level. We then populate these schemas and evaluate text-to-SQL systems on the same natural-language questions
across the resulting schema variants\footnote{The data generation scripts and datasets are available at: \url{https://github.com/umddb/text2sql-equivalent-schemas}.}. This gives us a systematic way to measure robustness to schema variation that arises from database design choices rather than from changes in domain or user intent.
Our work builds on recent work that studies the space of alternative relational representations for a common conceptual design~\cite{deshpande2025beyond}. That setting is a natural fit for the robustness
question we care about here, since it lets us generate many equivalent schema variants in a principled way. It also exposes a challenge that is easy to overlook in this setting: {\bf ground truth}. When two schemas differ, the correct SQL queries may also differ substantially in surface form even when they express the same intent. 
Hence, we focus on comparing the execution behavior across schema variants, i.e., we compare the results of the queries across the databases. We note that this provides, at best, a {\em lower bound on the inconsistency rates}
we can observe; an interesting direction of further research is to try to construct databases that can help better identify mismatches across the queries (analogous to work on killing SQL query mutants~\cite{shah2011generating}).

The main contributions of this paper are as follows:
\begin{itemize}
    \item We present a framework for generating and populating many equivalent relational schemas from a common E/R model by varying shredding choices.
    \item We use this framework to build text-to-SQL evaluation settings in which the natural-language questions remain fixed while the schema varies systematically.
    \item We evaluate several leading LLMs on these schema variants and show that they exhibit substantial inconsistencies (even if the E/R model is provided) across schemas that should, in principle, be equivalent.
\end{itemize}

More broadly, our results suggest that current text-to-SQL evaluations do not fully capture an important notion of robustness: invariance to alternative relational representations of the same underlying domain. We believe this
is a natural and practically important direction for evaluation, especially as LLM-based systems are increasingly used by non-experts who may not be sufficiently familiar with SQL. %over databases designed by different people and for different application needs.

Our work also suggests that the existing text-to-SQL systems could be made more robust by systematically incorporating schema variations into the process. For instance, one could create a few different equivalent databases
on the fly with different schemas (using an existing E/R diagram or reverse-engineering one~\cite{chiang1994reverse}), and generate SQL queries for a given natural language prompt simultaneously across those databases. 
Differences across the results of those queries could be used to flag potential problems with the SQL queries that are generated. We are planning to investigate this and similar approaches in our future work.

\begin{figure*}[t]
    \centering
    \includegraphics[width=.95\textwidth]{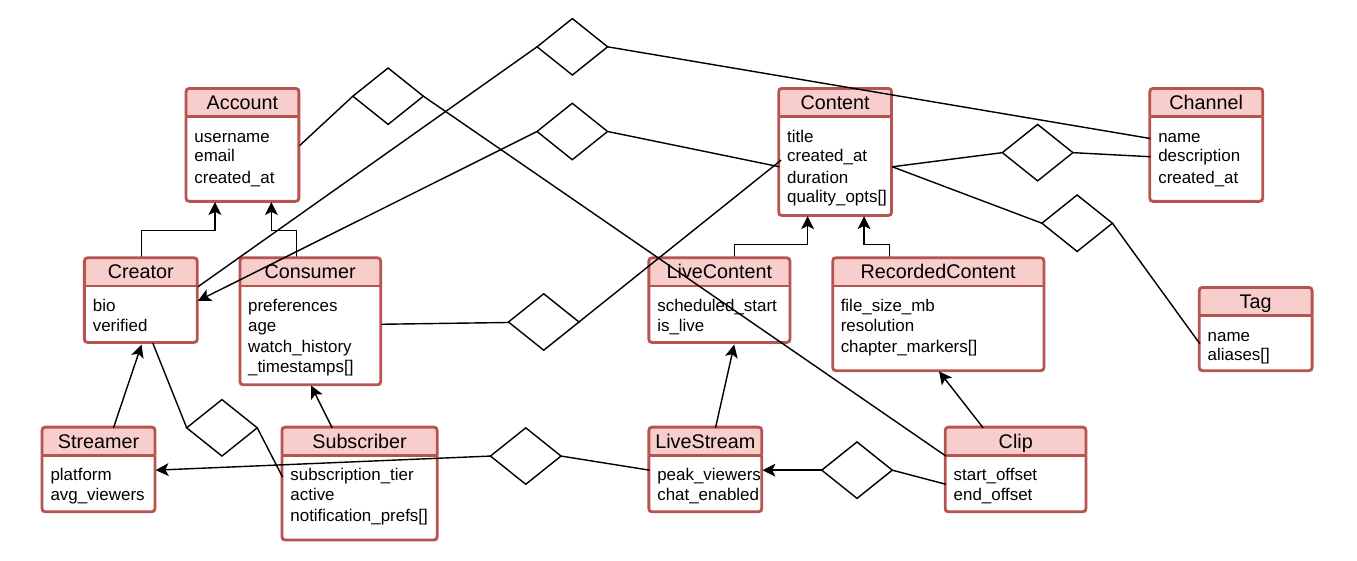}
    \vspace{-5pt}
    \caption{Social Media E/R diagram used as a running example. We omit relationship names for brevity.}
    \vspace{-5pt}
    \label{fig:social_er}
\end{figure*}

\section{Dataset Generation Framework}
\label{sec:framework}

In our framework, we treat an E/R specification as the semantic source of truth, and the relational schema as only one possible implementation of that specification. Starting from a single E/R diagram, we generate multiple
relational schemas that are \emph{conceptually equivalent}, populate them with instances derived from the same conceptual data, and then evaluate the same natural-language questions over all schema variants. This viewpoint is
closely aligned with the broader argument in prior work that the E/R model provides a more stable abstraction layer than the relational schema, and that choices such as how to represent inheritance, relationships, and
multi-valued attributes should be viewed as implementation decisions rather than user-visible semantics~\cite{deshpande2025beyond}.

Figure~\ref{fig:dataset_pipeline} shows the overall workflow, where we start with an E/R diagram and a small number of generation parameters, such as the desired scale of the data and distributions governing relationship cardinalities. From this, we first generate a conceptual instance of the E/R model, containing entity and relationship instances. We then apply a collection of shredding choices to translate the conceptual instance into multiple relational schemas and corresponding database instances. In parallel, we generate natural-language questions whose meaning is tied to the E/R model rather than to any one schema, and evaluate the same questions against each of the generated databases. 

The initial steps in this pipeline, specifically the generation of the E/R diagram and the natural-language questions, may be automated to generate new benchmarks. However, one could also read the schema from an existing database, and create schema variants from that (either through creating an E/R diagram first, or through applying rules directly to the relational schema).

\subsection{Running Example: Social Media Domain}
\label{subsec:running_example}

We use a social-media application as a running example throughout this section (Figure~\ref{fig:social_er}). The schema contains two nontrivial inheritance hierarchies, several multi-valued attributes, and a mix of many-to-one and many-to-many relationships, making it a good domain for illustrating the space of equivalent relational encodings.

Specifically, one inheritance hierarchy contains an \texttt{Account} entity, that is specialized into \texttt{CreatorAccount} and \texttt{ConsumerAccount}. Those two are further specialized into \texttt{Streamer} and
\texttt{Subscriber} respectively, with appropriate sets of attributes as shown. The other hierarchy is around \texttt{Content}, specialized into \texttt{LiveContent}, \texttt{LiveStream}, \texttt{RecordedContent}, and \texttt{Clip}. 
The schema also includes standalone entities such as \texttt{Channel} and \texttt{Tag}. Several attributes are multi-valued, including
\texttt{Tag.aliases}, \texttt{Content.quali- ty\_options}, \texttt{RecordedContent.chapter\_markers}, \texttt{ConsumerAc- count.watch\_history\_timestamps}, and \texttt{Subscriber.notificat- ion\_preferences}.

The relationship structure is equally rich. Every piece of content has a producer, so there is a many-to-one relationship from \texttt{Content} to \texttt{CreatorAccount}. Every live stream has a host, giving a many-to-one relationship from \texttt{LiveStream} to \texttt{Streamer}. Clips may be derived from live streams, giving another many-to-one relationship. The schema also contains many-to-many relationships indicating which content is published on which channels, which tags are attached to which content, which creators own which channels, which consumers viewed which content, which subscribers follow which creators, and which accounts shared which clips.
 %\hline
  %& Single & Class & Concrete \\ \hline
 %Tables & Single Table & Separate Tables & Separate Tables \\ \hline
 %Parent Columns & Stored in Single Table & Stored only in Parent Table & Stored in Both Parent \& Child Tables \\ \hline
 %Child-Specific Columns & NULL for Parent Records & Stored only in Child Table & Stored only in Child Table \\ 
 %\hline

\subsection{Generating Equivalent Relational Schemas} % from One E/R Model}
\label{subsec:equivalent_schemas}
Next, we discuss how we explore the family of equivalent relational schemas by systematically varying a small set of local design decisions.

One major source of variation is the {\bf representation of inheritance hierarchies}. In the social-media example, both \texttt{Account} and \texttt{Content} form multi-level subtype hierarchies. A schema designer could store all
members of a hierarchy in a single relation using a {\em discriminator} ({\em type}) attribute, often called {\bf single-table inheritance (STI)} pattern. 
Alternatively, the designer could place the common attributes in a root table and store subtype-specific attributes in separate subclass tables, corresponding to {\bf class-table inheritance (CTI)}. 
A third possibility is to create separate concrete tables for the leaf classes and duplicate inherited fields into those tables (called {\bf concrete table inheritance (CONCRETE)}). These are large structural changes but
preserve the semantics. A question about subscribers or streamers may require a simple filter predicate under one schema and a chain of joins under another.

\begin{table*}[t]
\centering
\small
\setlength{\tabcolsep}{5pt}
\renewcommand{\arraystretch}{1.15}
\begin{tabular}{p{2.7cm} p{4.2cm} p{3.8cm} p{4.8cm}}
\toprule
\textbf{Schema Variant} & \textbf{Inheritance Strategy} & \textbf{Multi-valued Attributes} & \textbf{Relationship Encoding} \\
\midrule

\texttt{full\_sti}
& Single-table inheritance for both \texttt{Account} and \texttt{Content} hierarchies; subtype membership represented using discriminator columns
& Prefer nested representation (arrays / JSONB) for attributes such as \texttt{aliases}, \texttt{quality\_options}, and \texttt{chapter\_markers}
& Many-to-one relationships represented as foreign keys; many-to-many relationships via junction tables \\

\texttt{full\_cti}
& Class-table inheritance throughout; each superclass/subclass gets its own relation with keys %along the hierarchy
& Fully normalized into separate child tables
& Many-to-one relationships represented as foreign keys; many-to-many relationships via junction tables \\

\texttt{full\_concrete}
& Concrete-table inheritance for leaf classes; inherited attributes duplicated into concrete subtype tables
& Fully normalized into separate child tables
& Relationship endpoints attached to concrete tables when possible; many-to-many relationships via junction tables \\

\texttt{full\_normalized}
& Class-table inheritance throughout
& Fully normalized into separate relations for every multi-valued attribute
& Explicit relationship tables used aggressively, including some many-to-one relationships \\%5that could otherwise be folded into foreign keys \\

\texttt{concrete\_all\_tables}
& Concrete-table inheritance for concrete subclasses, with root entities retained when needed %for shared references
& Fully normalized into separate relations
& Explicit tables for all many-to-many relationships and selected many-to-one relationships to maximize uniformity \\

\texttt{hybrid\_roots\_merged}
& Root and first-level hierarchy nodes merged; leaf subtypes such as \texttt{Streamer}, \texttt{Subscriber}, \texttt{LiveStream}, and \texttt{Clip} remain separate
& Mixed: frequently queried attributes normalized; smaller collections kept nested
& Foreign keys for many-to-one relationships; standard junction tables for many-to-many relationships \\

\texttt{hybrid\_leaves\_merged}
& Root entities represented separately, but leaf classes merged into parent subtype tables where possible
& Mixed: arrays / JSONB for compact attributes, normalized tables for larger collections
& Mostly foreign-key based, with junction tables for many-to-many relationships \\

\texttt{mixed\_sti\_cti}
& \texttt{Account} hierarchy uses STI, while \texttt{Content} hierarchy uses CTI
& Mixed normalized and nested representation
& Standard foreign keys for many-to-one relationships and junction tables for many-to-many relationships \\

\texttt{sti\_junction\_fks}
& STI for both major hierarchies
& Fully normalized for all multi-valued attributes
& Explicit foreign keys and junction tables, minimizing nested columns despite STI \\

\texttt{kitchen\_sink}
& Mix of STI, CTI, and concrete-style encodings across the two hierarchies
& Mixed representation chosen attribute-by-attribute
& Mixed encoding, including both foreign-key style and explicit tables \\ %relationship-table style realizations \\
\bottomrule
\end{tabular}
\vspace{10pt}
\caption{Representative schema variants generated from the same social-media E/R model. All ten schemas are semantically equivalent: they represent the same entities, relationships, and attribute values, but differ in how those concepts are shredded into relational structures.}
\label{tab:schema_variants}
\end{table*}

To make this concrete, consider the \texttt{Account} hierarchy. Under a single-table encoding (STI), a single \texttt{Account} table might contain columns for all root and subtype attributes, with a type discriminator indicating
whether a row is a generic account, creator, consumer, streamer, or subscriber. Under a class-table encoding (CTI), there may instead be separate relations such as \texttt{Account}, \texttt{CreatorAccount}, \texttt{ConsumerAccount},
        \texttt{Streamer}, and \texttt{Subscriber}, with the leaf types reconstructed by joining upward along the hierarchy. Under a concrete-table encoding (CONCRETE), \texttt{Streamer} and \texttt{Subscriber} may each carry their
        inherited attributes directly, without requiring the root relation to answer leaf-specific queries. %All three alternatives encode the same set of conceptual objects, but they present radically different relational surfaces to a text-to-SQL model.

A second major source of variation is the {\bf handling of multi-valued attributes}, which may be normalized into separate tables (with associated performance penalty due to additional joins needed), or 
may be stored in nested columns such as PostgreSQL arrays or JSONB values. %The CIDR paper argues that multi-valued attributes are naturally part of the E/R abstraction and need not be forced immediately into first normal form from the user’s perspective.  That observation is particularly useful here. In our benchmark generator, a multi-valued attribute may be represented in normalized form using a separate table, or it may be stored in a nested column such as a PostgreSQL array or JSONB value. 
For example, \texttt{Tag.aliases} may be represented by a relation \texttt{TagAlias(tag\_id, alias)}, or it may appear as an array-valued attribute inside \texttt{Tag}. Likewise, \texttt{Content.quality\_options} may become a relation \texttt{ContentQualityOption(content\_id, quality\_option)} or remain inside the \texttt{Content} table as a nested value. Both choices are semantically equivalent, but they lead to very different SQL. A query asking for clips that support ``1080p'' may compile to a join against a child table in one schema and to an array-containment predicate in another.

{\bf Relationships} introduce a third source of variation. For many-to-one relationships, the most common encoding is a \textbf{foreign key} on the many side, but this is not the only option. The producer relationship from
\texttt{Content} to \texttt{CreatorAccount}, for instance, can be encoded as a \texttt{producer\_id} attribute inside \texttt{Content}, or as an explicit relation such as \texttt{ProducedBy(content\_id, creator\_id)}.
Similarly, the \texttt{derived\_from} relationship between \texttt{Clip} and \texttt{LiveStream} can be represented either as a foreign key in \texttt{Clip} or as a standalone table. Many-to-many relationships almost always
induce junction tables in a fully normalized schema, but even there there is freedom in naming, key choices, and interaction with surrounding denormalization decisions. For example, if inheritance is encoded concretely, the
relationship \texttt{Subscriber-subscribes-to-CreatorAccount} may be realized against a subtype table in one schema and against a root table with type filtering in another.

In our framework, we first identify all such representational decision points by analyzing the provided E/R diagram. We then construct complete relational schemas by making different choices for those. In principle, this can
produce an exponential number of schemas, and we seek a representative set of variants that are both structurally distinct and realistic.

For the social-media domain, we generate ten schemas (Table~\ref{tab:schema_variants}). Some of these are ``extreme'' schemas that apply a single strategy consistently across the whole design. For instance, one schema may use class-table inheritance
for every hierarchy and normalize every multi-valued attribute into a separate relation. Another may use single-table inheritance wherever possible and preserve multi-valued attributes using nested types. These extremes are
useful because they maximize contrast and reveal whether models are sensitive to broad representational shifts. These also represent common strategies used in practice.

Other schemas are hybrid, to capture the fact real databases rarely adhere to a single design principle everywhere. A designer might choose class-table inheritance for the \texttt{Account} hierarchy but single-table inheritance for
\texttt{Content}; or might normalize watch histories and subscriptions, while keeping quality options or aliases in array-valued columns. Accordingly, we also include mixed schemas such as one in which the roots of both
hierarchies are merged but leaf classes remain separate, another in which leaf classes are merged while the roots stay normalized, and another in which inheritance is handled by STI but many-to-many relationships and
multi-valued attributes are represented by explicit junction tables. %These hybrid schemas are important because they resemble the kinds of compromises found in practice.

\subsection{Data and Query Generation}
\label{subsec:conceptual_data}
\label{subsec:nl_questions}

The conceptual instance generator operates directly over the E/R model. It first creates entity objects, assigns subtype memberships, instantiates multi-valued attributes, and then populates relationships. In the social-media domain, this means generating accounts, creators, consumers, streamers, subscribers, content objects, channels, and tags; assigning values such as verification status, average viewers, subscription tiers, or chapter markers; and then linking them through relationships such as production, hosting, publication, viewing, subscription, and sharing.

The generator is parameterized by scale and skew. We can control, for example, the total number of accounts, the fraction of accounts that are creators, the fraction of creators that are streamers, the number of channels owned per creator, the average number of tags attached to each content item, or the length distributions for arrays such as watch histories or quality options. These parameters allow us to create multiple benchmark sizes without altering the logical structure of the domain.

Once the conceptual instance has been created, each schema variant is populated by shredding the same objects and relationships into the target relations. The identities of conceptual objects are preserved across variants, so
that a particular clip, tag, or subscriber corresponds to the same underlying entity in every database. What changes is only the representation: a fact that appears as a nullable field in one schema may be split across multiple
tables in another, or may be nested inside an array or JSON structure in a third. This design gives us a clean controlled setting: the data is fixed, the question is fixed, and only the schema changes.

The set of natural language questions for testing purposes can either be generated manually (by a human), or could be auto-generated using an LLM. Since users usually reason about the entities and relationships in the data
when formulating their questions, in some ways, the questions are posed against the E/R model rather than any specific schema, and as a result, the same natural language question can be meaningfully posed to any of the schema
variants. 
For the social-media domain, some examples include: asking for streamers that hosted popular live streams, channels that publish content from verified creators, clips derived from streams hosted on a certain platform, or
subscribers who follow creators owning multiple channels. We use a relatively simple set of questions for our initial evaluation, since we could already see significant consistency issues across the schema variants; however, more complex questions can be easily added.

For the social media dataset, the questions used for evaluation were:
\begin{enumerate}
    \item Who are the 50 creators that own the largest number of channels?
    \item For each creator, how many total pieces of content have they produced?
    \item Who are the 50 streamers that have hosted the most live streams?
    \item How many clips were derived from each live stream?
    \item Who are the top 50 creators that have the largest number of subscribers?
    \item For each subscriber, how many different creators are they subscribed to?
    \item Which 50 channels have published the most content items?
    \item How many clips has each account shared?
    \item What are the 50 tags that are used on the largest number of content items?
    \item For each creator, how many times has content they produced been viewed by consumer accounts
    \item Who are the top 50 creators that produce content that is published on the greatest number of distinct channels?
    \item How many live streams hosted by each streamer have at least one derived clip?
    \item Which subscribers follow creators who own more than one channel?
    \item For each channel, how many live streams and how many clips have been published on it?
    \item Who are the top 50 creators that have produced content tagged with 3 or more tags?
\end{enumerate}

These questions were automatically generated using an LLM, with minor edits to make them less ambiguous. See Appendix~\ref{app:retail} for the questions used for the other dataset.

\subsection{Why These Variations Matter for LLMs}
\label{subsec:why_it_is_hard}

The social-media example makes clear why these schema variants can be difficult for current models to handle. A question such as \emph{``Which verified creators own channels that publish clips tagged with aliases of `gaming'?''}
is conceptually simple, but the data needed to answer it may be arranged across different sets of tables and connected in different ways. In one schema, \texttt{verified} may be a nullable field in a large \texttt{Account} table, \texttt{Clip} may have its own table, and tag aliases may require an unnest or JSON operator. In another schema, the same query may require joining \texttt{Account}, \texttt{CreatorAccount}, \texttt{RecordedContent}, \texttt{Clip}, \texttt{ContentTag}, and \texttt{TagAlias}. In yet another, some of these joins may disappear while others are replaced by discriminator predicates.

These differences matter because text-to-SQL models often rely heavily on surface correspondences between question words and schema tokens. Equivalent schemas disturb those correspondences without changing the intended meaning. Inheritance introduces ambiguity about whether a concept such as \emph{subscriber} or \emph{streamer} should map to a table, a subtype predicate, or a join path. Multi-valued attributes change not only where the relevant information is found but also what SQL operators are needed to access it. Relationship reification changes whether a connection is represented as a foreign key or as an explicit table. Our benchmark is designed precisely to test whether models can reason through such representational changes or whether their behavior is brittle.

\subsection{Using the Framework for Evaluation}
\label{subsec:framework_eval}

Suppose an E/R model yields schema variants $S_1, \dots, S_k$, and let $q_1, \dots, q_m$ be the natural-language questions. We evaluate each text-to-SQL model on every pair
$(q_i, S_j)$ using the respective APIs (the specific models used are listed in the evaluation section). 
For each schema-question combination, we generated the following prompt,
   replacing \textbf{schema} with the full contents of a PostgreSQL dump file containing \textit{CREATE TABLE} statements for the schema and \textbf{question} with the natural language question: \\[10pt] % we were asking for a SQL query for:
\texttt{You are the world's best natural language to SQL translator. You will be given two things: a schema for a PostgreSQL database and a natural language question. Given these two inputs, you should respond ONLY with a SQL
    statement that when run against the provided schema, returns the results that would answer the provided question. Again, you should ONLY answer with the SQL statement. Do not include any extra text before or after the SQL
        statement. When providing your query, you should provide the minimum number of columns to uniquely identify answers. For example, only return the ID of records that match your query and not other columns that are not
        necessary for unique identification unless the query explicitly specifies a list of columns. Return the minimal data necessary for both simple queries and aggregations. \\ The database schema is: \textbf{schema}. \\ The question is: \textbf{question}.}\\[3pt]

As we discuss later, the list of attributes that were selected (i.e., the \texttt{select} clause) varied widely in the generated SQL statements. To focus on the inconsistencies arising out of schema issues, we force the models to only return IDs. The rate of
mismatch was too high otherwise. Additionally, without the specification to only retrieve the ID fields, the LLMs sometimes omitted it altogether, making it hard to check for correctness or compare answers.

For each generated SQL query, we record the query, whether it executes successfully, and the resulting answer. Because the same conceptual instance underlies all schemas, these answers are directly comparable across variants.

This structure enables several forms of analysis. We can measure standard execution accuracy within each schema, but more importantly we can check whether the model’s behavior is stable across equivalent schemas. For example, we
can compare every pair of schemas and compute how often the model returns equivalent answers for the same question, or how often it fails on one schema after succeeding on
another. In the experimental section, we summarize these
cross-schema patterns using {\em pairwise heatmaps}, which make it easy to see which schema design choices induce the largest drops in consistency.

\subsection{Providing E/R-Diagram Context}
\label{sec:more_er}
A natural question is whether the performance of the text-to-SQL systems would be improved if the original E/R diagram is provided as part of the context (we note that this
is rarely available in practice). For one of the datasets (retail), we ran a second set of experiments by modifying the prompt: \\[5pt]
\texttt{ %You are the world's best natural language to SQL translator. You will be given the following: a description of an Entity-Relationship diagram for a domain, a schema and corresponding description for a PostgreSQL database that is derived from this ER diagram,and a natural language question with some context. Given these inputs, you should respond ONLY with a SQL statement that when run against the provided schema, returns the results that would answer the provided question. Again, you should ONLY answer with the SQL statement. Do not include any extra text before or after the SQL statement. \\ 
The Entity-Relationship diagram is: \textbf{ER-Diagram Context}. \\ The schema is: \textbf{schema}. \\ The associated context for the schema is: \textbf{schema description}. \\ The question is: \textbf{question}.}\\[3pt]
In addition to providing the list of entities (and their attributes), inheritance hierarchy, and relationships, we also provide a text description of the choices made in creating the relational schema (i.e., whether we used single-table inheritance, etc).

\section{Preliminary Evaluation}
In this section, we report the results from our experimental evaluation across four leading large language models (including Gemini, ChatGPT, Claude, and DeepSeek) across two datasets: social media and retail (Appendix~\ref{app:retail}). The specific models used are noted later.

\subsection{Evaluation Process}
For each natural language question, we generate SQL queries from each schema and execute them in PostgreSQL. %To do this, we provide the LLM with context of the schema and the natural language question. We specifically ask to return only the ID of the resulting columns, which allows us to have a strict number of columns in our return data.
To ensure consistent comparisons across schemas, we normalize the results as follows:
\begin{itemize}
    \item Rows are stored as a list of tuples
    \item Rows are sorted to remove ordering differences
    \item All values are converted to strings for uniform comparison
\end{itemize}
This normalization ensures that comparisons focus only on the actual result data returned, not formatting or ordering differences. 

We note that all of the above are important challenges for text-to-SQL systems in practice. Combined with the fact that we only test against a single instance of the data, the results we discuss below constitute a best-case scenario for the text-to-SQL systems.

\subsection{Pairwise Comparison Metrics}
\subsubsection{LLM Overview Metric}
For each pair of schemas $(i, j)$ and large language model $m$, we compute:
\[
\text{PSS}(i, j, m) =
\begin{cases}
\sum_{q=1}^{N} \mathbf{1}\big( R_{i,q,m} = R_{j,q,m} \big), & \text{if } i \neq j \\
\sum_{q=1}^{N} \mathbf{1}\big( R_{i,q,m} = R_{j,q,m} \big) - E_{i,m}, & \text{if } i = j
\end{cases}
\]

where:
\begin{itemize}
    \item $R_{i,q,m}$ = result of schema $i$ on question $q$ using the SQL query generated by model $m$
    \item $N$ = total number of questions (15)
     \item $E_{i,m}$ is the number of queries for schema $i$ and model $m$ that resulted in invalid SQL or execution errors.
\end{itemize}

Each pairwise schema score represents the number of matched queries between each pair of two schemas. We compute this score for every pair of schemas, resulting in a matrix where each row and column correspond to each schema and each entry represents their agreement out of 15.

\subsubsection{Pairwise Comparison by Question}
Additionally, for each question $q$ and large language model $m$, we compute the following scores:
\begin{itemize}
    \item \textit{QuestionAgreementScore} - the number of schema pairs for which an LLM returned valid SQL queries for both schemas and the queries resulted in equivalent result sets
\[
QAS(q, m) = \sum_{i,j=1, i < j}^{N} \mathbf{1}\big(R_{i,q,m} \ne \emptyset \land R_{j,q,m} \ne \emptyset \land R_{i,q,m} = R_{j,q,m} \big)
\]

    \item \textit{QuestionDisagreementScore} - the number of schema pairs for which an LLM returned valid SQL queries for both schemas, but the queries resulted in inequivalent result sets
\[
QDS(q, m) = \sum_{i,j=1, i < j}^{N} \mathbf{1}\big(R_{i,q,m} \ne \emptyset \land R_{j,q,m} \ne \emptyset \land R_{i,q,m} \ne R_{j,q,m} \big)
\]
    \item \textit{QuestionMissingScore} - the number of schema pairs for which an LLM returned an invalid SQL query for either one or both of the schemas
\end{itemize}
\[
QMS(q, m) = \sum_{i,j=1, i < j}^{N} \mathbf{1}\big(R_{i,q,m} = \emptyset \lor R_{j,q,m} = \emptyset)
\]

We note that: $QAS(q,m)+QDS(q,m)+QMS(q,m)=\binom{N}{2}$ where $N$ is the number of schemas. %We compute these three scores for each question and LLM, resulting in a breakdown of which questions LLMs were good at consistently generating correct SQL queries for and which ones caused issues.

\subsection{Overall Statistics}
\begin{figure}[t]
    \centering
    
    \begin{subfigure}[b]{0.47\linewidth}
        \centering
        \includegraphics[width=\linewidth,trim={3cm 1.5cm 3cm 0.5cm}]{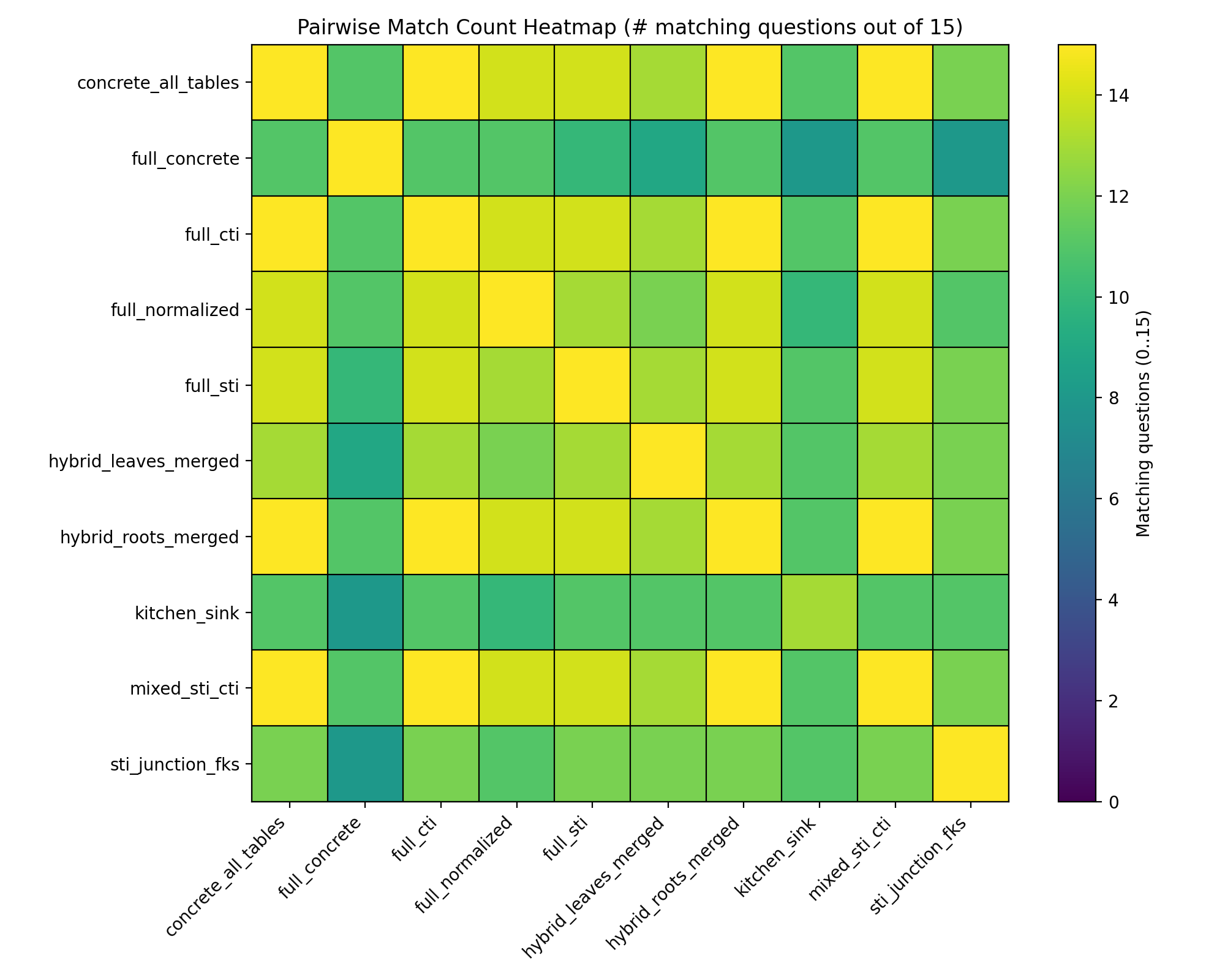}
        \caption{Gemini 3 Flash}
    \end{subfigure}
    \hfill
    \begin{subfigure}[b]{0.47\linewidth}
        \centering
        \includegraphics[width=\linewidth,trim={3cm 1.5cm 3cm 0.5cm}]{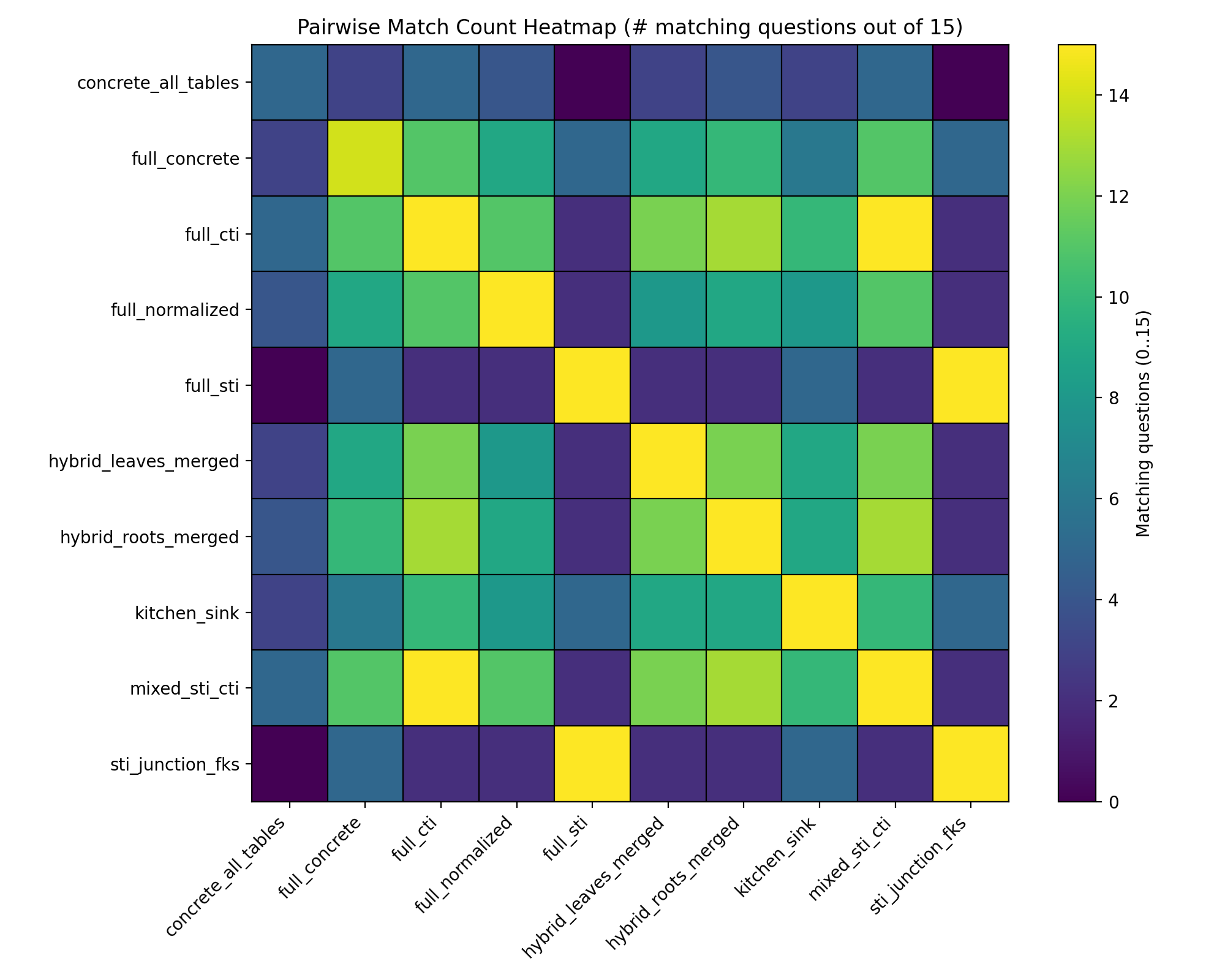}
        \caption{ChatGPT 5.2}
    \end{subfigure}

    \vspace{0.5em}

    \begin{subfigure}[b]{0.47\linewidth}
        \centering
        \includegraphics[width=\linewidth,trim={3cm 1.5cm 3cm 0.5cm}]{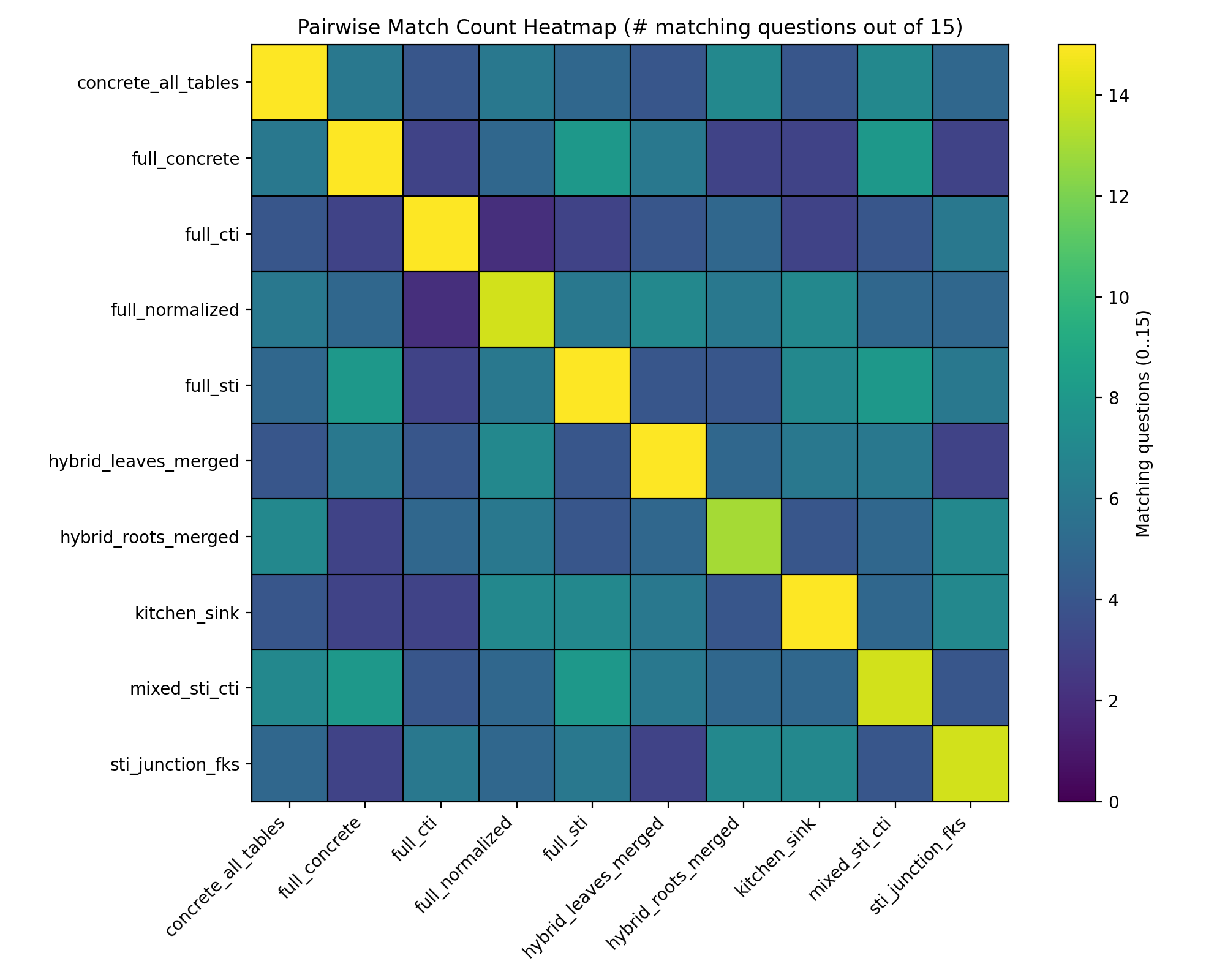}
        \caption{DeepSeek V3.2}
    \end{subfigure}
    \hfill
    \begin{subfigure}[b]{0.47\linewidth}
        \centering
        \includegraphics[width=\linewidth,trim={3cm 1.5cm 3cm 0.5cm}]{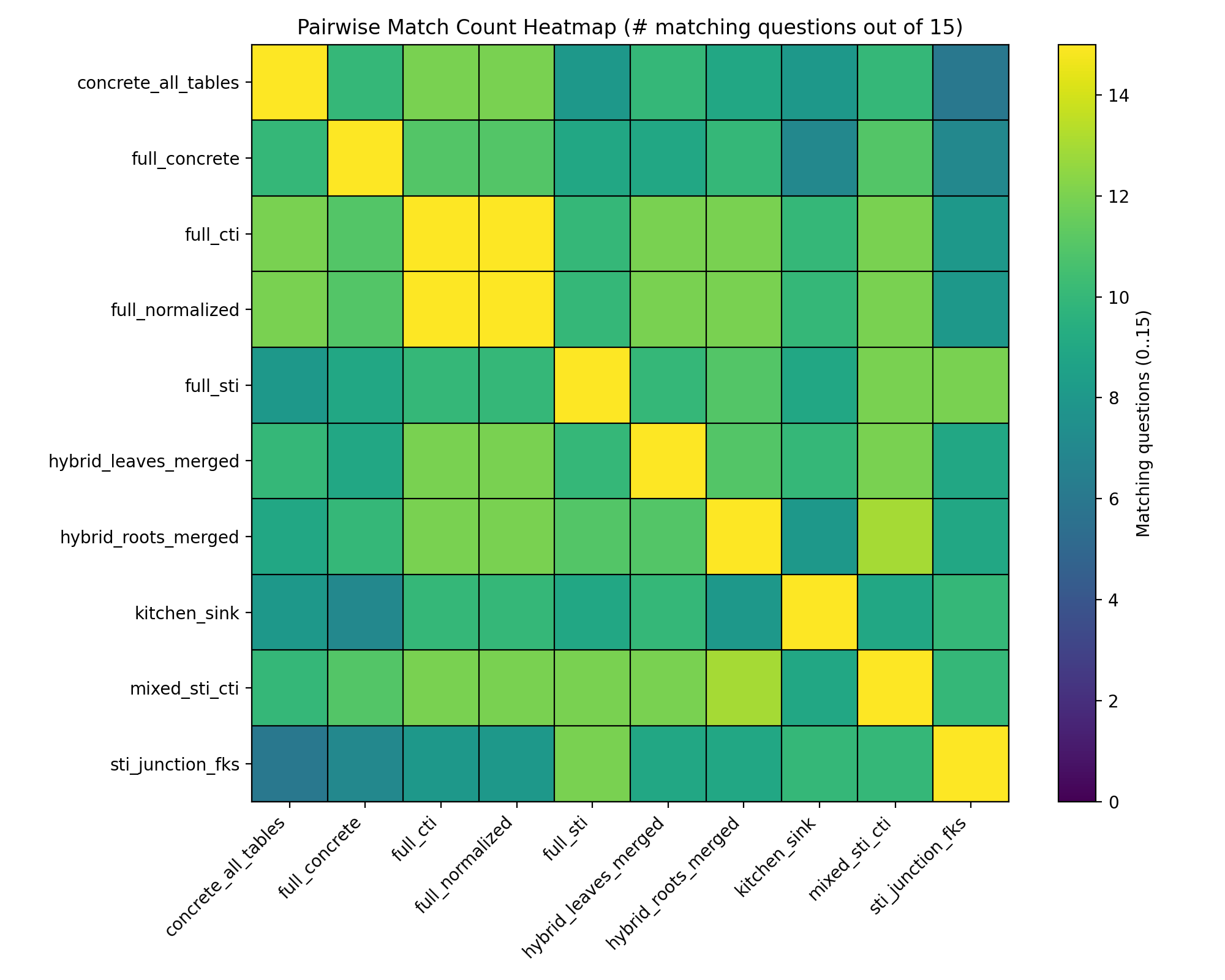}
        \caption{Claude Sonnet 4.6}
    \end{subfigure}

    \caption{Pairwise agreement heatmaps for all models for the \textbf{media dataset}. Lighter values indicate higher agreement. 
        Excluding the diagonal entries, the average agreements were 81.6\% (Gemini), 43.62\% (GPT), 33.85\% (DeepSeek) and 67.8\% (Claude).}
    \label{fig:media_heatmaps}
\end{figure}

\begin{figure}[h]
    \centering
    
    \begin{subfigure}[b]{0.47\linewidth}
        \centering
        \includegraphics[width=\linewidth,trim={3cm 1.5cm 3cm 0.5cm}]{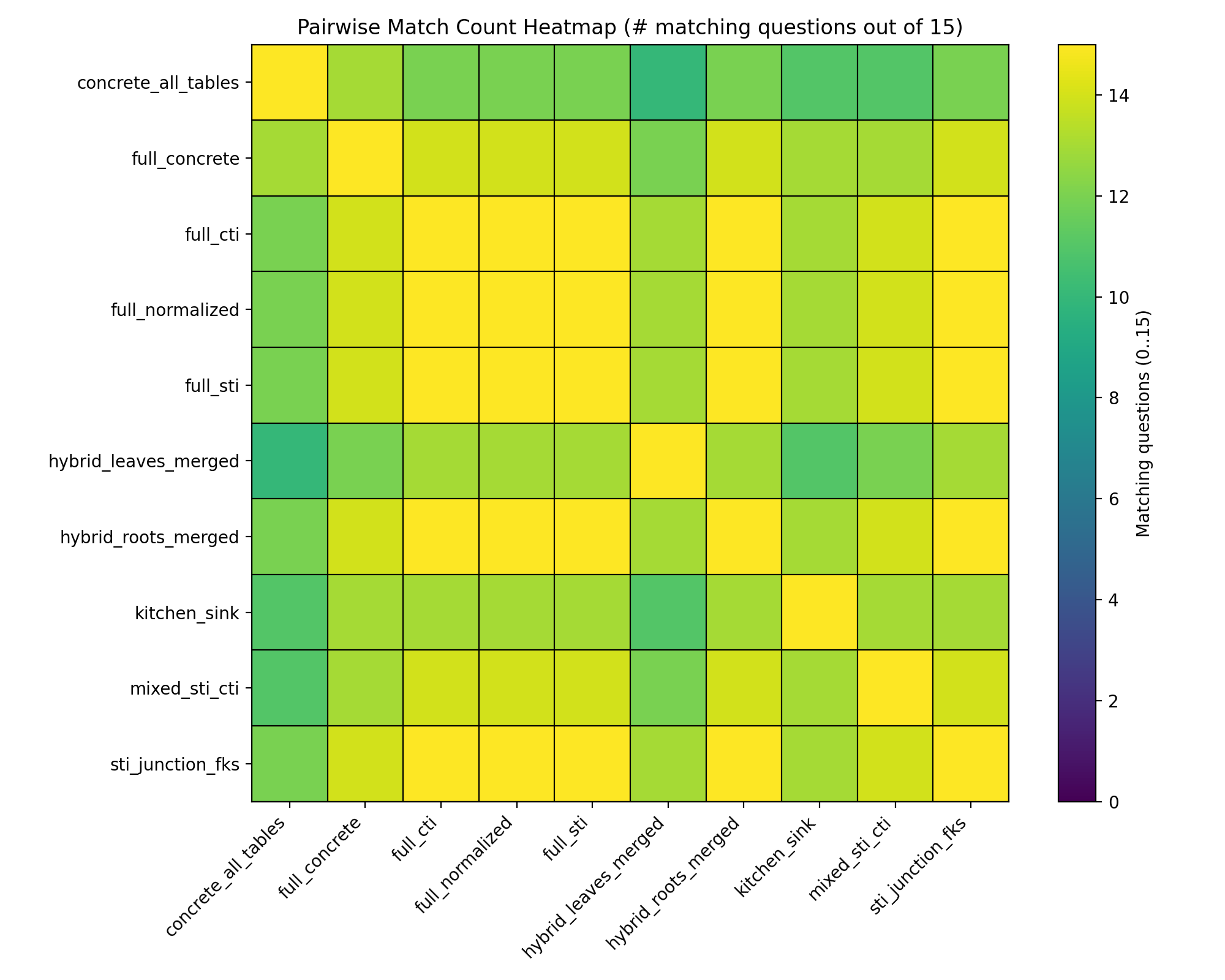}
        \caption{Gemini 3.1 Pro Preview}
    \end{subfigure}
    \hfill
    \begin{subfigure}[b]{0.47\linewidth}
        \centering
        \includegraphics[width=\linewidth,trim={3cm 1.5cm 3cm 0.5cm}]{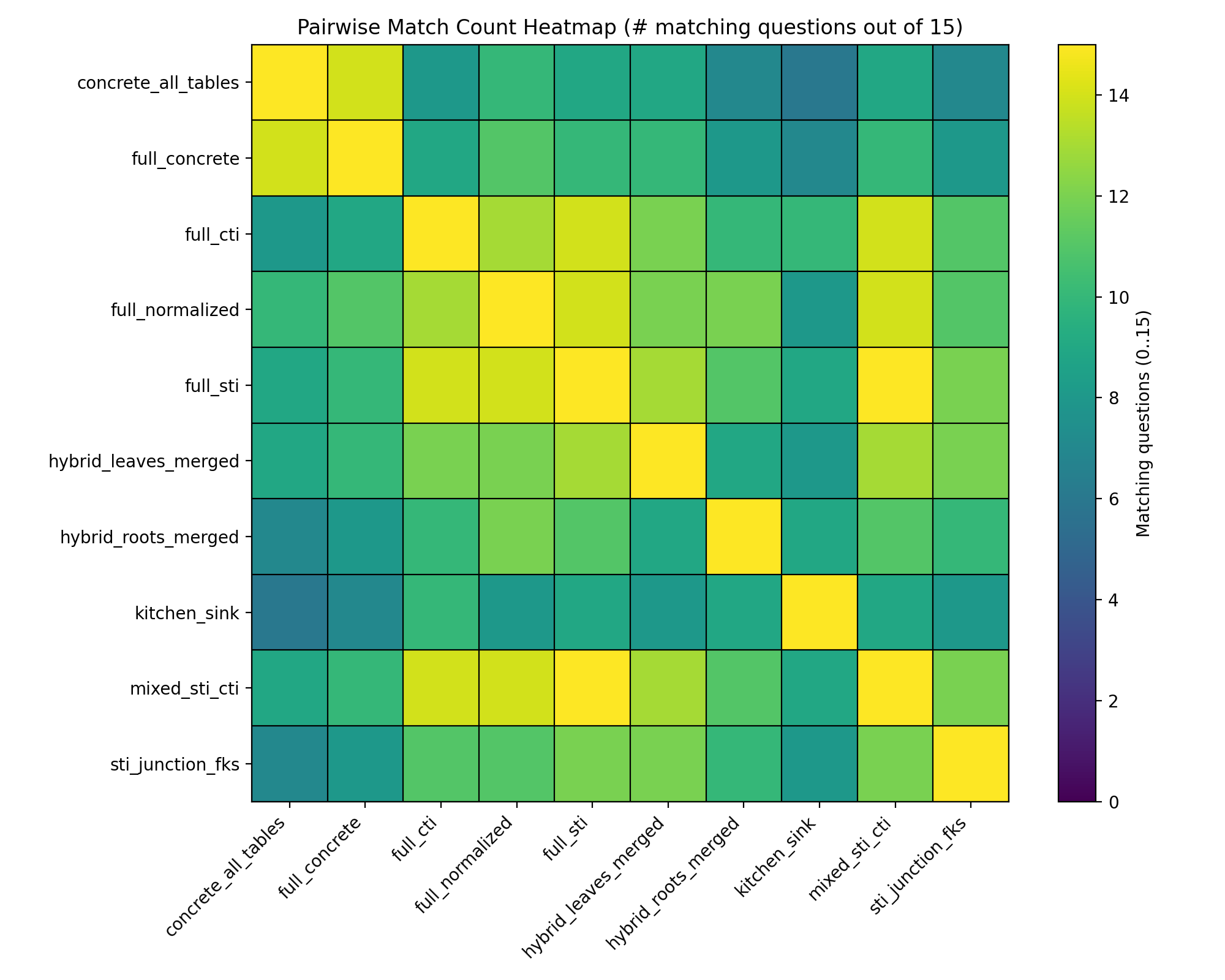}
        \caption{ChatGPT 5.4}
    \end{subfigure}

    \vspace{0.5em}

    \begin{subfigure}[b]{0.47\linewidth}
        \centering
        \includegraphics[width=\linewidth,trim={3cm 1.5cm 3cm 0.5cm}]{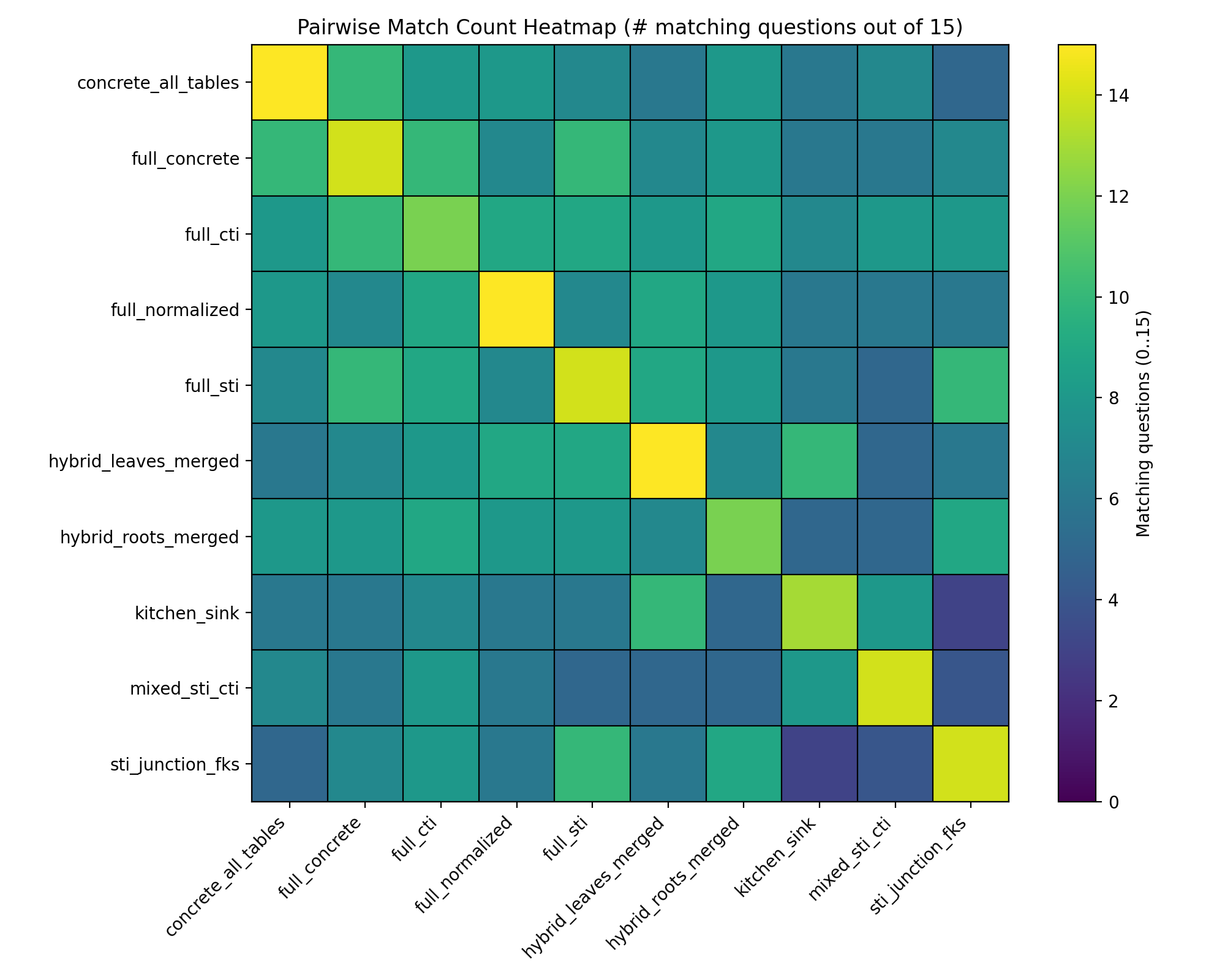}
        \caption{DeepSeek V4 Pro}
    \end{subfigure}
    \hfill
    \begin{subfigure}[b]{0.47\linewidth}
        \centering
        \includegraphics[width=\linewidth,trim={3cm 1.5cm 3cm 0.5cm}]{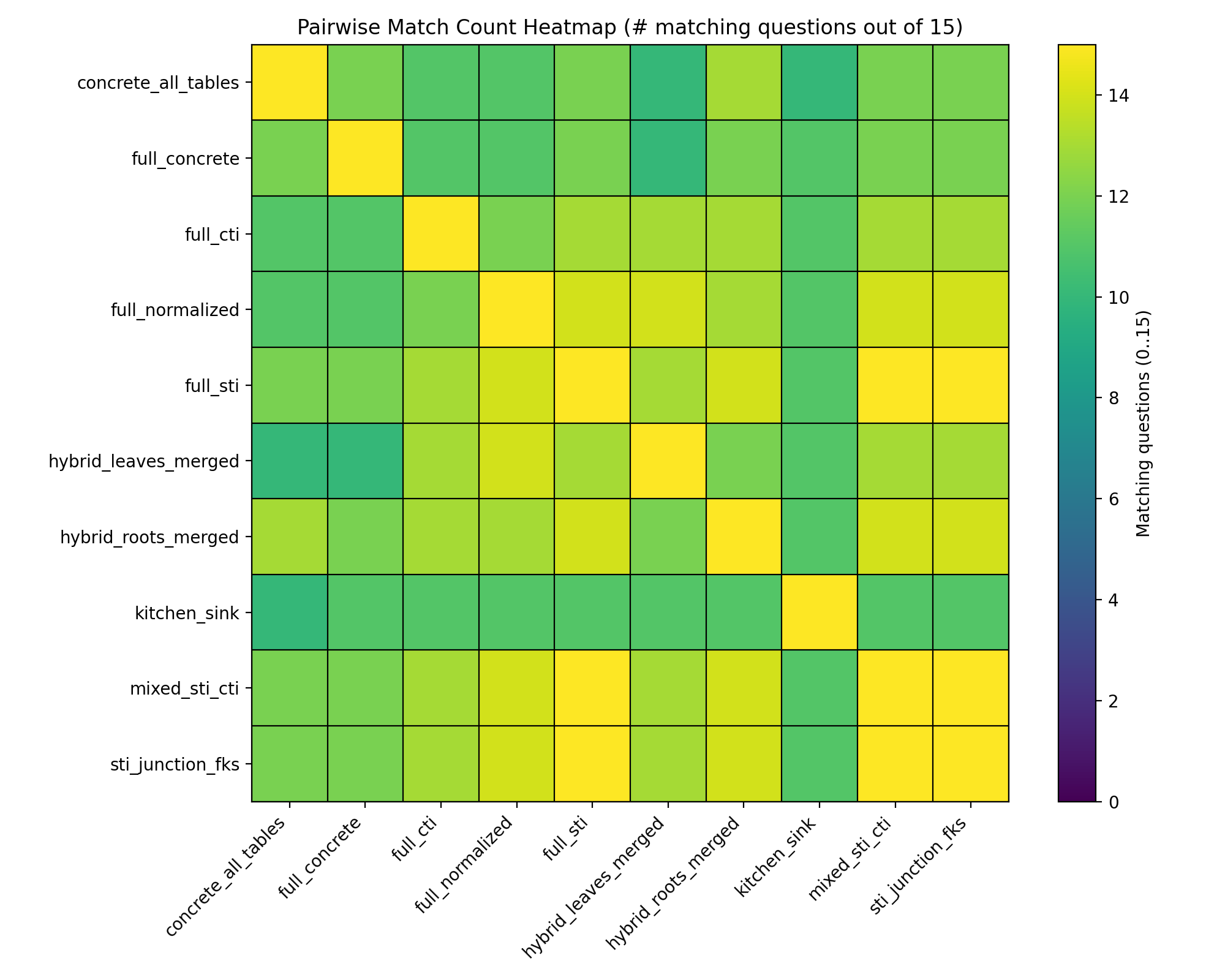}
        \caption{Claude Opus 4.7}
    \end{subfigure}

    \caption{Pairwise agreement heatmaps for all models for the \textbf{retail dataset}. Lighter values indicate higher agreement.
        Excluding the diagonal entries, the average agreements were 88.7\% (Gemini), 69.33\% (GPT), 47.4\% (DeepSeek) and 82.2\% (Claude). The higher agreement is likely because of 
            a less ambiguous set of questions as well as use of newer models.}
    \label{fig:retail_heatmaps}
\end{figure}

We visualize the pairwise agreement between the schemas using heatmaps, where the colors represent values from 0 (purple) to 15 (yellow).
As shown in Figures~\ref{fig:media_heatmaps} and~\ref{fig:retail_heatmaps}, Gemini achieves the highest overall agreement across schemas,
       followed by Claude and ChatGPT, while DeepSeek shows the lowest agreement. Note that the diagonal line remains consistently bright in all heatmaps, confirming that each schema agrees with itself (as it should). The only
       instance of imperfect agreement there occurs when a query throws an error when run in PostgreSQL. %However, the contrast between diagonal and off-diagonal cells highlights the extent to which schema variation impacts model consistency. These visual patterns demonstrate that disagreement is not random, but instead due to differences in schema.

In both the media and retail datasets, agreement varies significantly depending on the pair of schemas being compared. Certain schema pairs consistently produce high agreement, indicating that they are easier for models to understand. In contrast, other schema pairs exhibit substantial disagreement, reflecting differences in how relationships and entities are represented.

\noindent{\bf Media:} In the media dataset (Figure~\ref{fig:media_heatmaps}), Gemini 3 exhibits large regions of high agreement (bright yellow and green), especially among schemas like \texttt{full\_cti}, and \texttt{full\_normalized}. This
indicates that Gemini produces highly consistent queries across structurally similar schema. However, schemas such as \texttt{hybrid\_leaves\_merged} and \texttt{kitchen\_sink} have lower agreement rates on average.

ChatGPT 5 still has clusters of high agreement, but the heatmap is more fragmented with patches of medium and low agreement. The rows and columns corresponding to \texttt{full\_sti} and \texttt{concrete\_all\_tables} exhibit consistently darker values across many schema pairings, indicating that these schema have the most disagreement. In contrast, several hybrid schema, such as \texttt{hybrid\_leaves\_merged} and \texttt{hybrid\_roots\_merged}, show relatively higher agreement.

DeepSeek’s heatmap is darker overall, with fewer bright regions and many schema pairs showing low consistency. The diagonal remains bright, as expected, but the off-diagonal values are consistently lower, reflecting lower agreement across schema. Of the four models, DeepSeek tended to have the lowest agreement and the most syntactically incorrect SQL queries.

Claude Sonnet 4.6 displays a more structured pattern than DeepSeek but is less uniform than Gemini. Its heatmap shows clear bands of agreement, with strong consistency among mid-range schema such as \texttt{full\_cti}, \texttt{full\_normalized}, and \texttt{mixed\_sti\_cti}. However, the lowest agreement is concentrated in the corners of the heatmap, particularly when comparing structurally extreme schema such as \texttt{concrete\_all\_tables} and \texttt{sti\_junction\_fks}.

\vspace{5pt}
\noindent{\bf Retail:} The retail dataset (Figure~\ref{fig:retail_heatmaps}) shows overall higher agreements. This is attributable in part to a less ambiguous set of questions (see Appendix~\ref{app:retail}) as well as use of
newer models. Even then, we see significant disagrement rates, especially with GPT-5.4 and DeepSeek V4 Pro.

\subsection{Agreement by Question (Media Dataset)}
Next, we examine the models' performance on each NL question using the $QAS$, $QDS$, and $QMS$ scores described earlier on. Since each LLM generated 10 queries for each NL question, one query for each schema, there are $\binom{10}{2}=45$ possible pairs of queries by the same LLM on the same NL question. The agreement by question analysis reveals that LLM consistency varies significantly based on the complexity of the query.

Questions such as Q1 and Q4–Q9, which involve relatively straightforward aggregations or counting over a single relationship (e.g., counting channels, subscribers, or tags), exhibit higher agreement than the other questions for all models while for Gemini, they exhibit near perfect agreement across the 45 pairings. These queries usually require simple GROUP BY operations or minimal joins, making them less sensitive to how the schema is structured. DeepSeek, while its results exhibited mostly disagreement, had perfect agreement on Q6, which was a simple aggregation question over the subscriber-creator relationship, further showing that simple questions exhibit higher agreement. On the other hand, questions such as Q3, Q12, Q14, and Q15 show significantly higher disagreement. These queries require more complex reasoning over multiple relationships, such as linking live streams to derived clips (Q12), combining multiple entity types such as streams and clips (Q14), or enforcing complex conditions across joins such as “at least one derived clip” or “tagged with 3 or more” (Q12, Q15). As a result, models often fail to correctly reconstruct the necessary join paths or apply conditions at the appropriate stage of the query. Overall, these results demonstrate that LLMs perform reliably on simpler aggregation questions but struggle with questions that require a deeper relational understanding of the schema.

Interestingly, while Gemini showed the highest level of agreement in its SQL queries, Claude was the LLM that most consistently returned valid SQL with no invalid generated queries. On the other hand, Gemini and DeepSeek both produced a small portion of invalid statements and ChatGPT generated the most invalid SQL statements of the 4 models.

\begin{figure}[t]
    \centering
        \includegraphics[width=\linewidth]{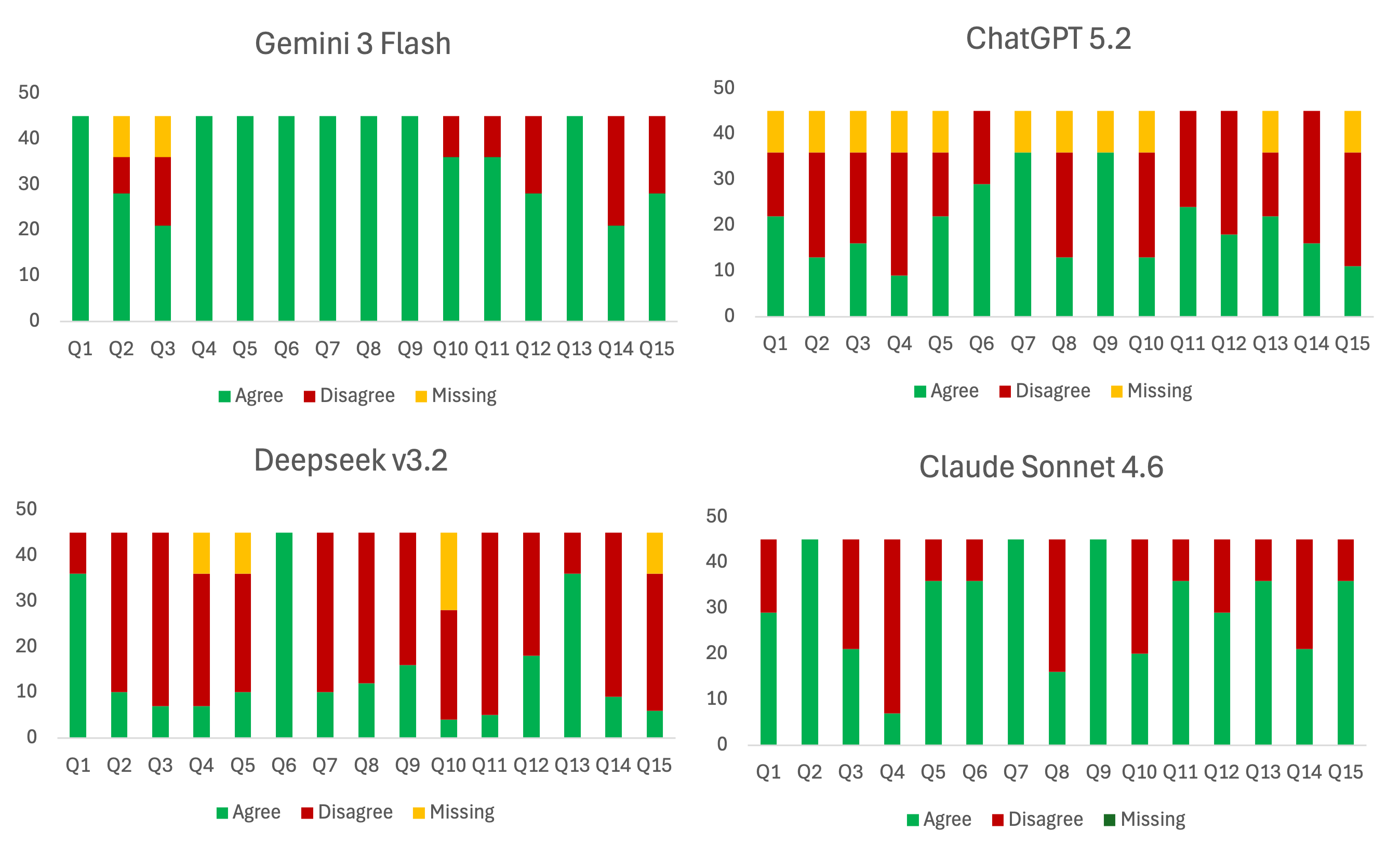}
    \vspace{-15pt}
    \caption{Agreement by question across schemas for each model on the {\bf media} dataset.}
    \vspace{-5pt}
    \label{fig:heatmaps_all}
\end{figure}

\subsection{Schema-Based Mismatches}

We observe several mismatches that are directly caused by differences in schema design.

\subsubsection{Representing Relationships in a Separate Table}
An example of mismatch caused by the schema differences occurs in Claude's responses to Q4, which asks for the number of clips derived from each live stream. In the \texttt{concrete\_all\_tables} schema, the query operates over a table (\texttt{clips\_derived\_from\_ref}) that only contains entries for live streams with at least one associated clip. Because of this, the aggregation over this table inherently excludes all live streams that have zero clips, since they do not appear in the table at all. In contrast, the \texttt{full\_cti} schema requires joining the \texttt{live\_streams} table with the \texttt{clips} table using a LEFT JOIN, which preserves all live streams and assigns a count of zero where no matching clips exist. 

This difference in schema design leads to different result sets: one includes only live streams that have had clips derived from them, while the other represents the full domain of live streams. The LLM fails to account for this
distinction, generating queries that are locally valid for each schema but semantically inconsistent. This highlights a key limitation in current text-to-SQL capabilities, where models do not fully adapt their query logic to
account for {\em how absence of relationships is encoded in different schema designs}.

A similar issue is found in Gemini's responses to Q3 for the \texttt{full\_concrete} and \texttt{hybrid\_leaves\_merged} schemas. Q3 asks for the IDs of the streamers that have hosted the most live streams. Since the \texttt{full\_concrete} schema has a separate \texttt{live\_streams} table, the list of returned host\_ids is accurate. On the other hand, for the \texttt{hybrid\_leaves\_merged} schema, the \texttt{live\_content} table holds all the LiveStream records. Since LiveContent records have \texttt{host\_id} set to NULL, the result for the \texttt{hybrid\_leaves\_merged} query says that the ID of the creator that has hosted the most live streams is NULL with 2000 hosted live streams, an obvious error.

\subsubsection{Type Discrimination in Multi-Entity Tables}
Another example of mismatch appears in Claude's responses to Q4 when comparing the \texttt{concrete\_all\_tables} and \texttt{hybrid\_leaves\_merged} schemas. In the \texttt{hybrid\_leaves\_merged} schema, all live content related entities are stored in a single table (\texttt{live\_content}), with a type attribute identifying between general live content and specifically live streams. The generated query
aggregates over the entire table without filtering specifically for live streams, resulting in a significantly larger result set that includes records corresponding to non-stream and stream objects. On the other hand, the \texttt{concrete\_all\_tables} schema isolates live streams into a dedicated structure, ensuring that only valid live stream records are counted. This leads to approximately 2000 additional
records in the \texttt{hybrid\_leaves\_merged} result, because the LLM fails to correctly interpret the type constraint required in schema that merge multiple entity types into a single table. This indicates a broader challenge with hybrid schema, where semantic distinctions are encoded as attribute values rather than through structural separation, requiring more precise filtering logic that LLMs often do not include.

\subsubsection{Confusion from Schema Hierarchy Changes}

Another schema caused mismatch occurs in Q10 when comparing the \texttt{concrete\_all\_tables} and \texttt{kitchen\_sink} schemas. In \texttt{concrete\_all\_tables}, creators are represented through a dedicated \texttt{creator\_accounts} table. In contrast, the \texttt{kitchen\_sink} schema collapses parts of the account hierarchy, removing a separate \texttt{creator\_accounts} table and instead combining all attributes into the broader \texttt{accounts} table while keeping \texttt{streamers} as a subtype. The generated query incorrectly joins through \texttt{streamers} instead of filtering the \texttt{accounts} table, which restricts the result to only streamers and not creators who aren't streamers. As a result, the query returns only 1480 rows instead of 2843. This mismatch shows that when schema hierarchies are restructured, LLMs may confuse a subtype with the full entity set.

\subsection{Non-Schema-Based Mismatches}

We also observe errors that are not directly caused by schema differences.

\subsubsection{Data Based Errors}
A common class of errors across models consists of queries that return only zeros, empty result sets, or \texttt{NULL} values despite executing successfully. These failures are often due not to schema misinterpretation, but to limited awareness of the underlying data values and relationships. For example, since data values are not included in the prompt context, a model may generate filters that do not match the actual data distribution, such as using \texttt{"clip"} instead of the correct \texttt{"Clip"}. Similarly, improper join conditions or aggregation logic can produce \texttt{NULL} values when table relationships are handled incorrectly. Such errors show that even with a correctly interpreted schema, insufficient data awareness can still lead to invalid or misleading outputs.

\subsubsection{SQL Syntax Errors}
A non-schema-related failure appears in Q3 for the \texttt{kitchen\_sink} schema, where the generated query uses the invalid PostgreSQL expression \texttt{ORDER BY COUNT()} with no argument. The query therefore fails to execute and produces no output. This error reflects malformed SQL generation rather than any issue with the schema, illustrating that basic SQL correctness remains a separate source of failures even when the intended query structure is otherwise plausible.

\subsubsection{Logic Based Errors}
Another example of error is when the LLM returns two different responses for the same table structure as shown in DeepSeek's responses to Q13, which asks which subscribers follow creators that own more than one channel. In \texttt{concrete\_all\_tables} and \texttt{full\_concrete} schemas, the \texttt{subscribers},  \texttt{subscribers\_subscribes\_to}, and \texttt{creator\_accounts\_owns} tables have the same structure. However, while both schemas presented the model with the same structure, only the first query returned the correct results. The other query performed a three-way join between the tables and then only grouped by subscriber ID, which led to it returning all subscribers whose followed creators had more than one channel when combined rather than those whose followed creators included a single creator that had more than one channel.
These errors highlight limitations in the models' ability to correctly interpret query intent, independent of schema structure.

\subsection{Performance by Schema}
Overall, we observe that schemas with clear separation between entities tend to lead to higher agreement rates, while schemas that store multiple relationships or merge multiple entity types lead to significantly more mismatches.
Schemas such as \texttt{concrete\_all\_tables} and \texttt{full\_normalized}, which explicitly represent relationships through dedicated tables and foreign keys, are generally easier for models to reason with. These schemas make join paths explicit, reducing ambiguity in query generation.

In contrast, hybrid and denormalized schemas such as \texttt{hybrid\_leaves\_merged}, \texttt{hybrid\_roots\_merged}, and \texttt{kitchen\_sink} had lower agreement rates. These schema often merge multiple entity types into a
single table or encode relationships through implicit attributes. This requires the model to infer filtering conditions, which are frequently omitted, leading to overcounting, undercounting, or inclusion of irrelevant records. A possible resolution is providing more context in the prompt to the model. 

We find that the most challenging schemas are those that deviate from standard normalization patterns by either over-merging or over-abstracting relationships. These designs require more complex reasoning, which current LLMs struggle with. On the other hand, schemas that align more closely with conventional relational modeling principles tend to produce query results with higher agreement.

\subsection{Impact of Providing E/R Context}
\begin{figure}[t]
    \centering
    
    \begin{subfigure}[b]{0.47\linewidth}
        \centering
        \includegraphics[width=\linewidth,trim={3cm 1.5cm 3cm 0.5cm}]{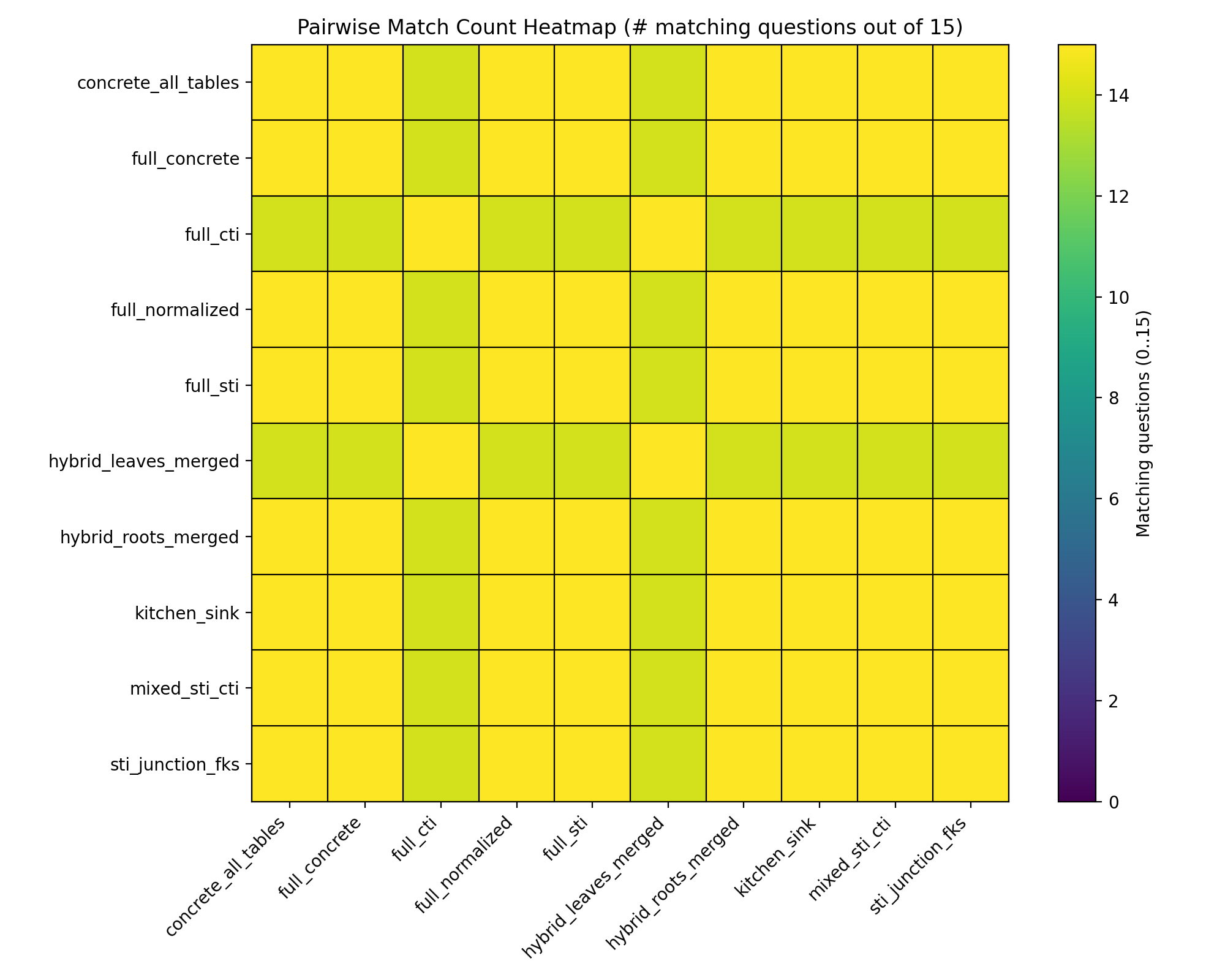}
        \caption{Gemini 3.1 Pro Preview}
    \end{subfigure}
    \hfill
    \begin{subfigure}[b]{0.47\linewidth}
        \centering
        \includegraphics[width=\linewidth,trim={3cm 1.5cm 3cm 0.5cm}]{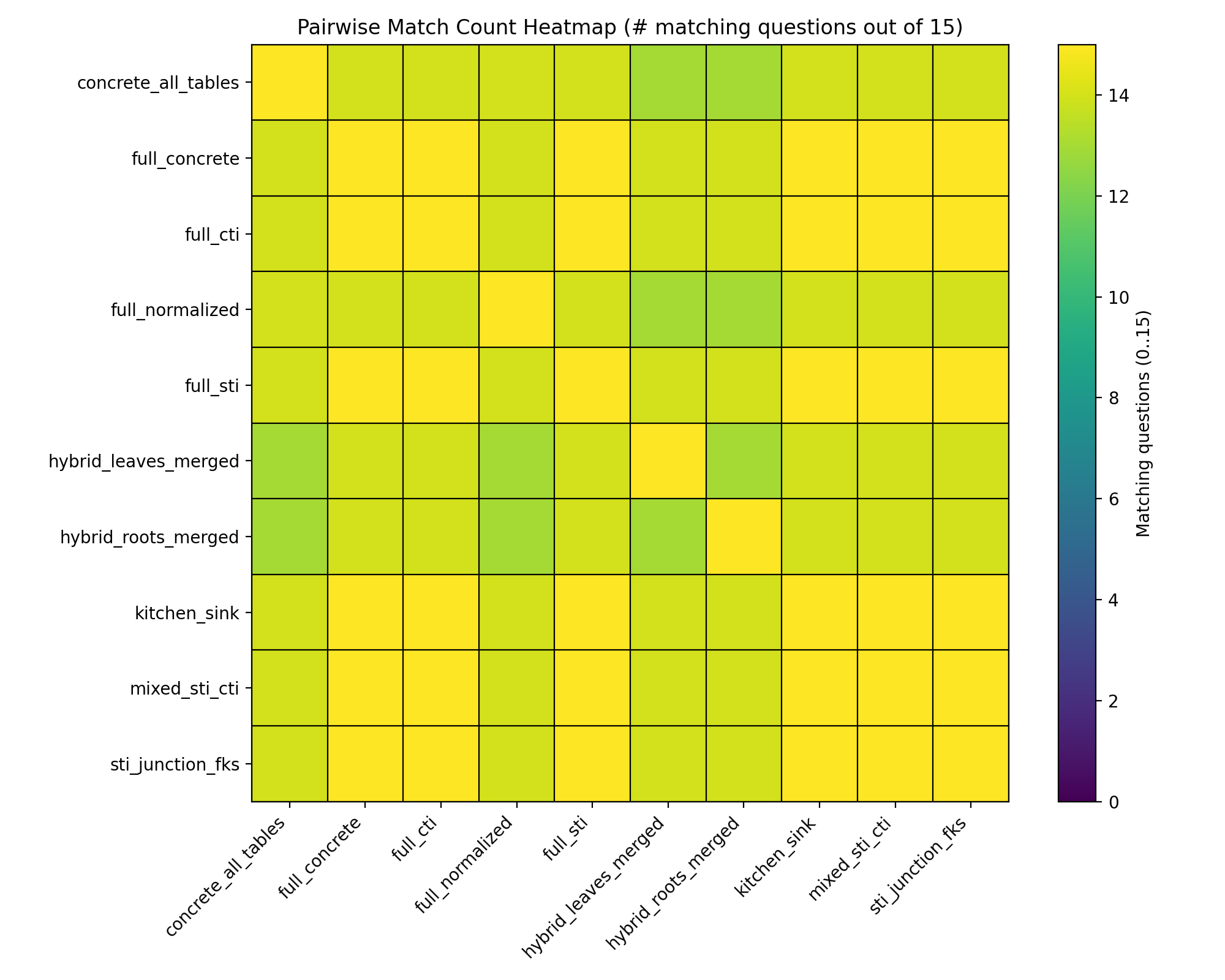}
        \caption{ChatGPT 5.4}
    \end{subfigure}

    \vspace{0.5em}

    \begin{subfigure}[b]{0.47\linewidth}
        \centering
        \includegraphics[width=\linewidth,trim={3cm 1.5cm 3cm 0.5cm}]{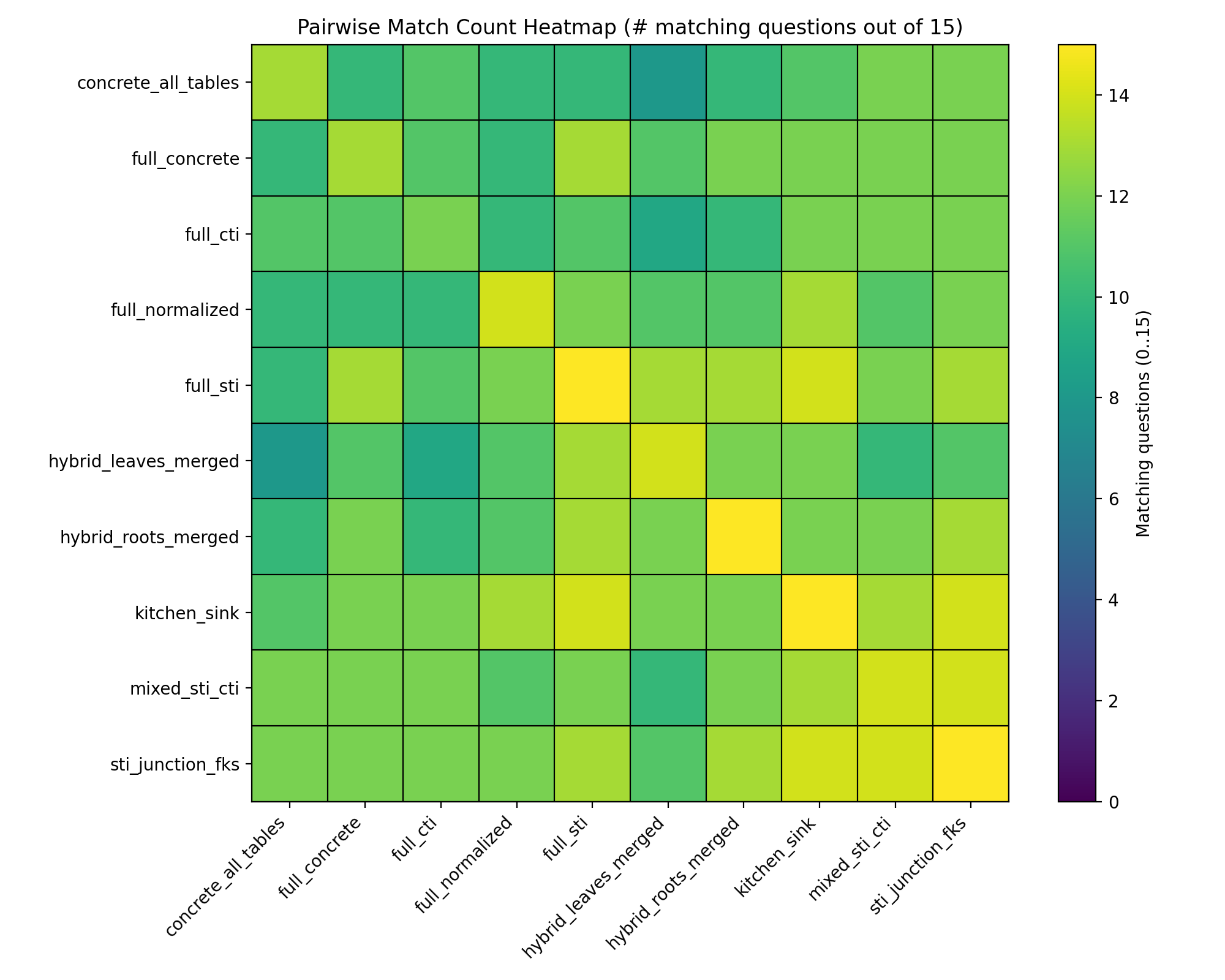}
        \caption{DeepSeek V4 Pro}
    \end{subfigure}
    \hfill
    \begin{subfigure}[b]{0.47\linewidth}
        \centering
        \includegraphics[width=\linewidth,trim={3cm 1.5cm 3cm 0.5cm}]{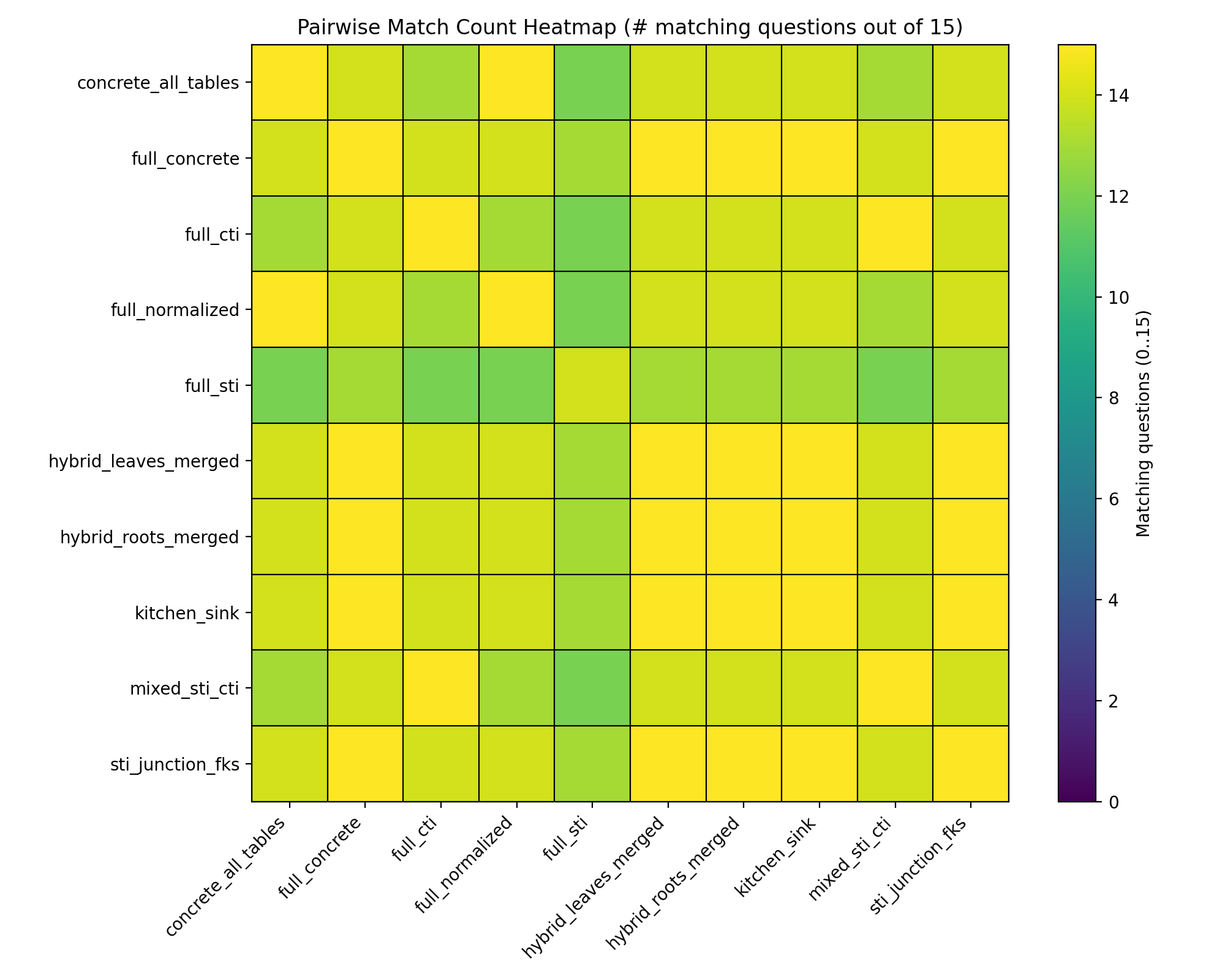}
        \caption{Claude Opus 4.7}
    \end{subfigure}

    \caption{Pairwise agreement heatmaps for all models for the \textbf{retail dataset} with additional ER \& schema context. Lighter values indicate higher agreement.}
    \label{fig:media_heatmaps_er_context}
\end{figure}

Figure~\ref{fig:media_heatmaps_er_context} shows the effect of adding the full E/R context described in Section~\ref{sec:more_er} for the retail dataset. The added domain and schema information substantially improves SQL-generation accuracy across all models. Nevertheless, the models still exhibit an overall disagreement rate of approximately 10\%. As noted earlier, this is a best-case setting for text-to-SQL evaluation, since correctness is tested against a single, relatively uniform dataset; disagreement rates are likely to be higher in practice, especially with messier data and more complex questions.

The improvement suggests that many earlier errors were due not simply to missing table relationships, but to missing domain-level information about what those relationships represented. For example, without being told that a query targets the \texttt{full\_sti} schema, a model may not infer the higher-level structure implied by the schema design, particularly when subclass relationships are encoded using single-table inheritance as discussed earlier. Providing the E/R context helps expose this structure and leads to more accurate queries.

Looking deeply at the 15 questions, 
we saw that Question 9 was the only question where all of the models had issues: ``List the 3 Apparel items that are paired with the greatest number of distinct Accessories items. Return the apparel product name, category, available sizes array, and the count of accessories it is paired with. Order by accessories count descending; break ties by product name ascending.'' The discrepancy is due to the LLMs performing
slightly different aggregations across the three-way table joins. Specifically, in the `full\_cti` schema, the apparel item with the most distinct accessories is ``Leather Swimwear'' with 5 distinct accessories. However, in the `concrete\_all\_tables` schema, the returned record for ``Leather Swimwear'' is only paired with 4 distinct accessories. Multi-table aggregation queries are fairly complex, so it is to be
expected that even an LLM that has been provided with a lot of context would struggle with these questions.

\section{Conclusion}
Our results show that current LLM-based text-to-SQL systems are not robust to schema variations. Across schemas
that are logically equivalent and populated from the same underlying E/R instance, we observe substantial
differences in both the generated SQL and the final answers. Taken together, the experimental results suggest
that these models remain sensitive to representational details such as normalization, inheritance mapping, and
relationship encoding, rather than reasoning reliably over the underlying conceptual structure. 
Our results also suggest that adding relevant E/R context can significantly improve the performance of these models.
More broadly, this points to a gap in how text-to-SQL systems are currently evaluated. Benchmarks built around a
single schema can mask an important notion of robustness that matters in practice, since real
applications often admit multiple plausible relational designs for the same domain. %Our framework addresses this by holding the data and questions fixed while varying only the schema, making it possible to measure consistency across equivalent representations instead of against just a single target query.

\vspace{4pt}
\noindent\textbf{Limitations:} 
Our study has several limitations. First, the question set is still relatively small, and therefore does not capture the full diversity of query patterns that arise even in moderate-sized schemas. Expanding the benchmark to
include more questions, more templates, and more complex reasoning patterns would provide a fuller picture of
model behavior. 
Second, our equivalence checks are based primarily on single-shot execution results. While practical, they do not capture interactive usage of these tools~\cite{huo2026birdinteract}; they also do not
have sufficient coverage and may classify queries as being equivalent when they are not. %all edge cases involving ordering, aggregation, and alternative but semantically equivalent outputs. 
      Third, our setup assumes perfectly equivalent schemas derived from a common E/R model; real-world schema evolution
is often messier, involving partial overlap, added or removed attributes, and drifting semantics. Finally, lack of ground truth restricts the ability to derive stronger conclusions about the performance of the models; however,
   this could be remedied relatively easily by constructing ground truth queries against the simplest schemas, or the E/R diagram itself.
   %Finally, our analysis focuses on correctness and consistency, rather than on other important properties such as efficiency, latency, and readability of the generated SQL.

\vspace{4pt}
\noindent\textbf{Future Work:} 
A natural next step is to scale up the benchmark along several dimensions: more schema families, more domains, and a much broader set of natural-language questions. It would also be valuable to study robustness at the level of
individual schema transformations, in order to identify which representational changes most consistently affect model behavior. On the modeling side, our results suggest the need for text-to-SQL systems that reason over
schema-invariant conceptual structure rather than over surface relational form alone, potentially through intermediate representations grounded in entities and relationships. More ambitiously, the framework introduced here could
also be used not just for evaluation, but for training: equivalent schema variants offer a principled way to generate additional supervision and to encourage models to generalize across multiple relational realizations of the same
underlying semantics. %Finally, incorporating execution cost, latency, and query quality into the evaluation would move this line of work closer to the requirements of real deployments.

\bibliographystyle{plain}
\bibliography{main}

\appendix

\section{Retail Dataset}
\label{app:retail}

\begin{figure*}[t]
    \centering
    \includegraphics[width=\textwidth]{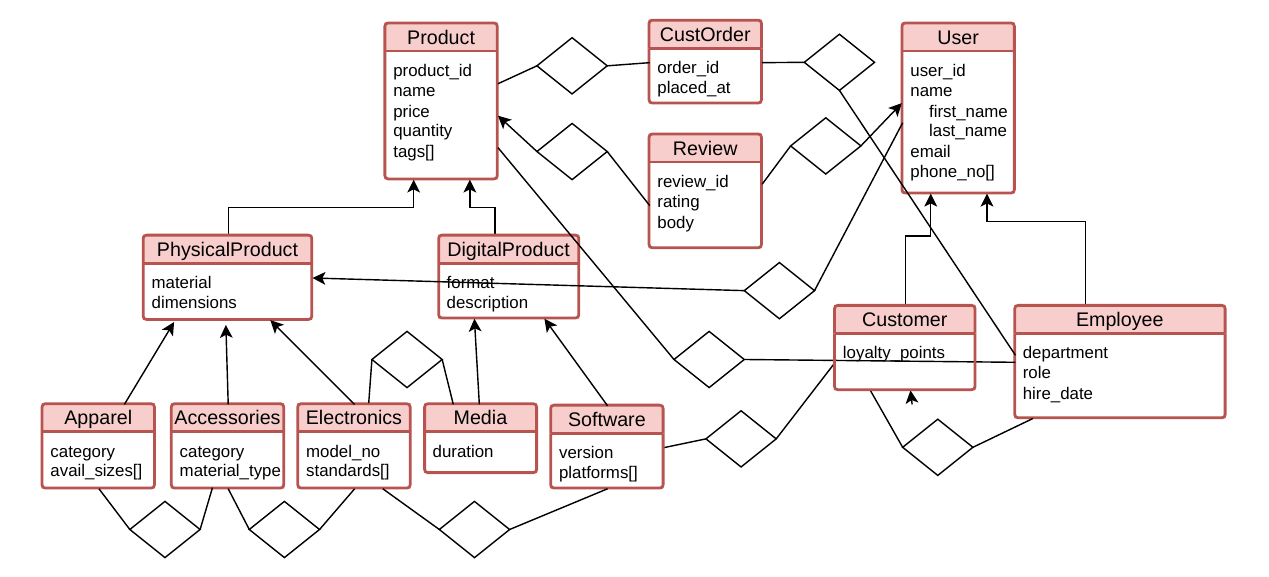}
    \caption{E/R Diagram for the Retail Dataset}
    \label{fig:retail_er}
\end{figure*}

\vspace{4pt}
\noindent
Figure~\ref{fig:retail_er} shows the E/R diagram for the retail dataset. The natural-language questions we use for this database are:

\begin{enumerate}
\item For each department, compute the total number of employees in that department and the average number of orders processed per employee, rounded to two decimal places. Include all departments that have at least one employee. Order by average orders processed descending; break ties by department name ascending.
\item For each department, report the number of employees in that department, the total number of distinct products recommended by those employees, and the total number of distinct orders processed by those employees. Return one row per department. Order by department name ascending.
\item For each of the 1,000 Electronics products, report the number of distinct Software products installable on it and the number of distinct Media products that can stream to it. Order by the sum of the two counts descending; break ties by product name ascending.
\item For each of the five concrete product types: Apparel, Accessories, Electronics, Media, and Software, report the total number of products of that type and the average price rounded to the nearest integer. Return one row per product type. Order by average price descending.
\item For every Accessories-Electronics pair, return the accessories product name, accessories category, accessories price, electronics product name, electronics model number, and electronics price. Order by accessories product name ascending, then by electronics product name ascending.
\item List every customer who has downloaded at least one Software product. Return the customer's first name, last name, email, loyalty points, and the count of distinct Software products they have downloaded. Order by download count descending; break ties by last name ascending.
\item List every customer who has placed at least 3 orders and written at least 2 reviews. Return the customer's first name, last name, email, order count, and review count. Order by order count descending, then review count descending, then last name ascending.
\item List every review that has a rating of exactly 5. Return the reviewer's first name, last name, the name of the reviewed product, and the review body text. Order by reviewer last name ascending, then reviewer first name ascending.
\item List the 3 Apparel items that are paired with the greatest number of distinct Accessories items. Return the apparel product name, category, available sizes array, and the count of accessories it is paired with. Order by accessories count descending; break ties by product name ascending.
\item List the 5 customers with the highest loyalty point totals among those who have written at least one review. Return the customer's first name, last name, email, loyalty points, and their total review count. Order by loyalty points descending; break ties by last name ascending.
\item List the 5 customers with the highest total spend. Define total spend as the sum of the prices of all products that appear in orders the customer has placed: for each order the customer has placed, add the price of every product in that order (a product appearing in two different orders of the same customer is counted twice, once per order). Return the customer's first name, last name, email, loyalty points, and total spend. Order by total spend descending; break ties by last name ascending.
\item List the 5 employees who have processed the highest number of distinct orders. Return the employee's full name, department, role, and the count of orders they have processed. Order by order count descending; break ties by employee name ascending.
\item List the 5 employees whose recommended products span the greatest number of distinct product types, where product type is one of: Apparel, Accessories, Electronics, Media, Software. Return the employee's full name, department, role, the count of distinct product types represented in their recommendations, and their total number of recommended products. Order by distinct product type count descending; break ties by total recommended products descending, then by employee name ascending.
\item List the 5 most expensive Electronics products by price. Return the product name, model number, price, and the list of supported standards. Order by price descending; break ties by product name ascending.
\item List the 5 Software products that are installable on the greatest number of distinct Electronics items. Return the software name, version, price, and compatible electronics count. Order by compatible electronics count descending; break ties by software name ascending.
\end{enumerate}

\end{document}